\documentclass{iopart}

\usepackage{iopams}	

\usepackage{latexsym}
\usepackage[dvips]{graphicx}
\usepackage{subfigure}
\usepackage{color}
\usepackage{float}

\usepackage[latin1]{inputenc} 
\usepackage[cyr]{aeguill}

\newcommand{\ket}[1]{|{#1}\rangle}
\newcommand{\bra}[1]{\langle{#1}|}
\newcommand{\ud}{\mathrm{d}}

\newcommand{\eqref}[1]{(\ref{#1})}

\begin{document}

\title{Adiabatically coupled systems and fractional monodromy}

\author{M S Hansen\dag, F Faure\ddag, B I Zhilinski\'i\S}

\ead{M.S.Hansen@mat.dtu.dk, \\ 
\hskip1cm  frederic.faure@ujf-grenoble.fr, \\
\hskip1cm zhilin@univ-littoral.fr}

\submitto{\JPA} 

 \begin{abstract}
We present a $1$-parameter family of systems with fractional 
monodromy and adiabatic separation of motion. We relate the 
presence of monodromy to a redistribution of states both in 
the quantum and semi-quantum spectrum. We show how the 
fractional monodromy arises from the non diagonal action 
of the dynamical symmetry of the system and  manifests itself  
as a generic property of an important subclass of adiabatically 
coupled systems.

\end{abstract}

\address{\dag\ Institute of Mathematics, Technical University of Denmark, 2800 Kgs. Lyngby, Denmark}
\address{\ddag\ Institut Fourier, Universit\'e Joseph Fourier, BP 74, 
  38402 Saint-Martin d'H\`eres,  Cedex, France}
\address{\S\ Université du Littoral, UMR du CNRS $8101$, $59140$ Dunkerque, France}

\section{Introduction}

Adiabatically coupled systems are systems with a slow and a 
fast motion in interaction. Such systems appear generically 
in different fields of physics, chemistry, biology. Examples 
of such systems are abundant: spin-precession in a slowly 
varying magnetic field, Foucault's pendulum, rovibrational 
or vibronic motion of molecules etc. Properties of such 
systems are therefore of a general interest.

The idealized physical simplification for such systems 
consists, for example,  in  representing the slow motion as being 
``infinitely'' slow in the adiabatic limit 
from the point of view of the fast 
degrees of freedom.  Conversely, on the slow time-scale 
the fast fluctuations are supposed to cancel out in such a way that 
the slow degrees of freedom see only the averaged fast motion. 

\begin{figure}[h!]
\begin{center}
    \input{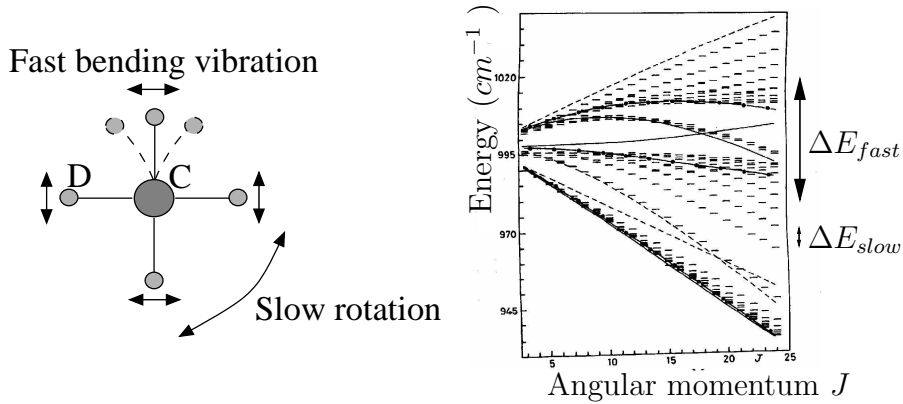}
\end{center}
\caption{Energy spectrum (shown right) for the molecule CD$_4$ 
(shown left) as a function of 
the angular momentum quantum number. The presence of
band structure is due to 
the fact that the vibrational motion is much faster than the 
rotational motion.}
\label{article:fig:spectrumExample}
\end{figure}

Due to  Heisenberg's uncertainty principle
\begin{equation}
    \Delta E\Delta \tau\sim\hbar \Leftrightarrow \Delta E\sim 
             \frac{\hbar}{\Delta\tau}=\hbar\omega,
\end{equation}
a separation of time-scales $\Delta \tau_{fast}\ll \Delta \tau_{slow}$
gives rise to $\Delta E_{fast}\gg \Delta E_{slow}$, i.e.  an energy spectrum with a 
structure in bands. \Fref{article:fig:spectrumExample} gives an example 
of the rovibrational energy spectrum of  the molecule
CD$_4$. Here energy levels are additionally classified by the value 
of the angular momentum $\boldsymbol J$ which is a strict integral
of motion. 
More precisely \fref{article:fig:spectrumExample} (right) shows the joint 
spectrum of two commuting operators representing the Hamiltonian and 
angular momentum of the molecular system. The three bands are due to 
three vibrational excited quantum states forming a fundamental polyad 
of the triply degenerate bending mode $\nu_4$ of CD$_4$. The internal 
structure of each band originates in the slower rotational motion of 
the entire  molecule.

Quantum joint spectrum shown in  
\fref{article:fig:spectrumExample} is calculated on the basis of 
effective Hamiltonian for $\nu_2/\nu_4$ dyad of CD$_4$ 
\cite{CanJPhys} and it is represented (according to \cite{pavlov88})
together with classical
energies of relative equilibria \cite{SIADS} which explain
the principal qualitative features of the corresponding
quantum band structure.

One generic feature of adiabatically coupled systems is a redistribution 
of energy levels between bands as some parameter varies. 
In \fref{article:fig:spectrumExample} the angular momentum plays the 
role of such a parameter. Another natural choice of parameter could be 
the magnitude of an external magnetic field \cite{cushman00},
vibrational polyad energy/quantum number \cite{SIADS}, etc.

Sometimes the corresponding classical system is integrable or 
can be approximated by integrable one by constructing the so-called
normal form \cite{arnold89,bates97}. In this 
case it is interesting to establish relations between such qualitative
feature of integrable approximation as Hamiltonian
monodromy and the phenomenon of the 
redistribution of energy levels between bands which is the
characteristic property of the initial adiabatically 
coupled system.

In this article we consider a simple $1$-parameter family of 
Hamiltonians which is a slight generalization of the well-known example of
spin-orbit coupling. This latter model has been the object of 
several studies \cite{pavlov88,sadovskii99,faure00,grondin} 
demonstrating the presence of integer monodromy 
for some interval of parameter values. 

We remind here, that the Hamiltonian 
monodromy is a generic property of 
classical integrable systems, intensively studied and popularized
by R. Cushman (see \cite{bates97}) and 
described in details 
by J.J. Duistermaat in $1980$ \cite{duistermaat80} In classical
dynamical systems with two degrees of freedom the Hamiltonian
monodromy can typically appear in one-parameter families
through Hamiltonian Hopf bifurcation \cite{duistermaat98}.
It was shown later that there is a correspondence between the 
appearance of monodromy within a one-parameter family of classical 
Hamiltonians and the redistribution of bands in the spectrum of the 
associated quantum problem \cite{sadovskii99,ngoc03}. 
The appearance of Hamiltonian monodromy
in classical system indicates also   
the presence of a  topological bifurcation
in a semi-quantum (Born-Oppenheimer) description \cite{faure01}.

Our model has \emph{fractional monodromy} which is a recent 
generalization of integer monodromy concept 
\cite{nekoroshev03,nekoroshev03b,KECushSad,KEbook}. 
This is the first example of a system with this property on a compact 
phase space  and we demonstrate how the change in monodromy type leads 
to a change in the redistribution pattern.


This article is a part of the ongoing study of global properties of 
integrable systems on one side 
\cite{broer02,ngoc99,zung96,zung02,nekoroshev03b} - 
especially in the context of molecular physics 
\cite{cushman00,cushman04,joyeux03,sadovskii99,SIADS,MolPhys,child,AnnPhys} 
- and adiabatically 
coupled systems   \cite{berry84a,panati03,faure00,faure02a} on another side.

\section{Presentation of model}

\subsection{Dynamical symmetry and Hamiltonian}
\label{article:symmetry}

Very often global properties of the dynamical model under study 
are due to the symmetry of  the physical problem under consideration. 
The model we study in this paper admits  a non-diagonal 
group action of $G=SO(2)$
\begin{eqnarray}  \label{article:groupaction}
SO(2)\times (S^2\times S^2)&\rightarrow 
                             S^2\times S^2 \\ \nonumber
(\phi;N_+,N_-,N_z,S_+,S_-,S_z) &\mapsto 
       (N_+e^{i\phi},N_-e^{-i\phi},N_z,S_+e^{2i\phi},S_-e^{-2i\phi},S_z)
\end{eqnarray}
on two coupled effective angular momenta 
$\boldsymbol N=(N_x,N_y,N_z),\boldsymbol S=(S_x,S_y,S_z)$ with fixed
$| \boldsymbol N|=\sqrt{N_x^2+N_y^2+N_z^2}$ and 
$| \boldsymbol S|=\sqrt{S_x^2+S_y^2+S_z^2}$. In  \eqref{article:groupaction}
$N_{\pm}=N_x\pm iN_y,S_{\pm}=S_x\pm iS_y$. The action 
defined by \eqref{article:groupaction}
can be considered  as initial data imposed by the  physical model.

As soon as the group action is given, a generic 
Hamiltonian can be constructed as a linear combination of 
polynomials invariant under  the group action \eqref{article:groupaction}. 
This leads  to a Hamiltonian which has an $SO(2)$ symmetry generated 
by $J_z=2S_z+N_z$, (i.e. $[H_{\lambda},J_z]=0$):

\begin{equation}
\label{article:hamiltonian}
    \fl H_{\lambda}=\frac{1-\lambda}{|\boldsymbol S|}S_z
      +\lambda\left( \frac{1}{|\boldsymbol S||\boldsymbol N|}S_zN_z
      +\frac{1}{2|\boldsymbol S||\boldsymbol N|^2}\left(N^2_-S_+
      +N^2_+S_-\right)\right)\textrm{, }0\leq\lambda\leq 1.
\end{equation}
Here $\lambda$ is a coupling parameter. It can be due to an 
external magnetic field, for example. 
The amplitudes $|\boldsymbol S|,|\boldsymbol N|$ are held fixed 
and we only consider the case $|\boldsymbol N|>2|\boldsymbol S|$. 
\footnote{A preliminary study of the case $|\boldsymbol N|
 <2|\boldsymbol S|$ has been initiated in \cite{hansen04}.} 

The $SO(2)$ symmetry generated by 
$J_z=2S_z+N_z$ rotates simultaneously  $\boldsymbol N$ and 
$\boldsymbol S$ about their respective $z$-axes. In \cite{sadovskii99} 
the $SO(2)$ action on the phase space $S^2\times S^2$ was diagonal but 
now the asymmetric appearance of $\boldsymbol N$ and $\boldsymbol S$ 
implies that while 
$\boldsymbol N$ is rotated by an angle $\phi$,  $\boldsymbol S$ is rotated 
by $2\phi$.

\subsection{Quantum description and semi-classical limit}

Conceptually it is more convenient to go from a quantum to a 
classical system and we begin by a presentation of the quantum 
system.\footnote{Several quantum systems may give rise to the 
same classical system. See e.g. \cite{zhilinskii01} for an example.}

$\boldsymbol N,\boldsymbol S$ are the angular momentum operators 
\cite{landau65} spanning an irreducible representation of 
$\mathfrak{su}(2)\times\mathfrak{su}(2)$ in a Hilbert space 
$\mathcal H=\mathcal H_N\otimes\mathcal H_S$ of dimension 
$(2N+1)(2S+1)$. $N,S$ are the respective angular momentum quantum numbers 
taking integer or half-integer values and 
$|\boldsymbol N|=\sqrt{N(N+1)},|\boldsymbol S|=\sqrt{S(S+1)}$.

The quantum dynamics are given by the Schr\"odinger equation 
(with $\hbar\equiv 1$)
\begin{equation}
    i\frac{\rmd}{\rmd t}\ket{\psi}=\hat H_{\lambda}\ket{\psi},
\end{equation}
where $\ket{\psi}$ is a vector in $\mathcal H$. To study the 
semi-classical limit of large quantum numbers $N,S\gg 1$ we introduce 
the normal symbol of $\hat H$ \cite{nakahara90,faure01}
\begin{equation}
\label{article:symbol}
    \bra{\boldsymbol N,\boldsymbol S}
     \hat H_{\lambda}\ket{\boldsymbol N,\boldsymbol S}=
                  H_{\lambda}+\Or(\hbar_{N,S}),
\end{equation}
which is a power series in $\hbar_N=1/(2N),\hbar_S=1/(2S)$. 
$\ket{\boldsymbol N,\boldsymbol S}$ are $SU(2)$ coherent states 
often used to study the semi-classical limit of angular momentum 
dynamics \cite{leboeuf91,kurchan89,zang90}.

Keeping only the first term of \eqref{article:symbol} we have a 
classical Hamiltonian, $H_{\lambda}$, which is the principal symbol 
of $\hat H_{\lambda}$. The dynamics is approximately described 
by \emph{classical} angular momenta $\boldsymbol N,\boldsymbol S$ 
moving according to Hamilton's equations of motion \cite{arnold89} 
\begin{eqnarray}
    \frac{\ud}{\ud t}\boldsymbol N
          &=&\hbar_N\partial_{\boldsymbol N}H_{\lambda}
                                 \wedge\boldsymbol N,\nonumber\\
    \frac{\ud}{\ud t}\boldsymbol S
             &=&\hbar_S\partial_{\boldsymbol S}H_{\lambda}
                 \wedge\boldsymbol S,\label{article:hamilton}
\end{eqnarray}
on the phase space which is topologically the direct product of
two two-dimensional spheres, $S^2\times S^2$. 
Putting $\hbar_{N,S}\rightarrow 0$ 
illustrates how the semi-classical limit is related to the 
limit of adiabatically slow motion. 
Under additional assumption $N\gg S$ giving $\hbar_N \ll \hbar_S$, the Hamilton's equations 
\eqref{article:hamilton} describe the dynamics of an adiabatically 
coupled system with the motion of $\boldsymbol N$ being much slower 
than that of $\boldsymbol S$.


\section{Classical description: Structure of the moment map}

\subsection{Second integral of motion}

The $SO(2)$ symmetry gives rise to a second integral of motion
\begin{equation}
\label{article:jz}
    J_z=2S_z+N_z,\qquad \{H_{\lambda},J_z\}_{S^2\times S^2}=0,
\end{equation}
which is the projection of the total angular momentum 
$\boldsymbol J=2\boldsymbol S+\boldsymbol N$ onto the $z$-axes. 
Together $H_{\lambda},J_z$ define a one-parameter family of 
integrable systems with two degrees of freedom.


\subsection{Reduction of symmetry, space of orbits}
\label{article:sec:orbitspace}

The symmetry of the system can be used to reduce the number of 
degrees of freedom. This is done by mapping each orbit of the 
$SO(2)$-action on $S^2\times S^2$ onto the $3$ dimensional space 
of orbits. As the group action is not transitive this is an 
example of so-called singular reduction \cite{bates97} based 
on the theory of invariants \cite{bates97,michel01,sadovskii99}. 

\begin{figure}[h!]
    \centering
	     \input{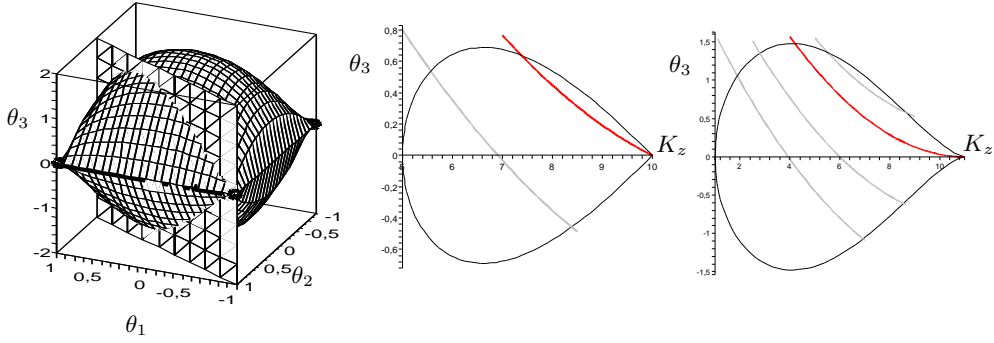}   	    	 
    \caption{Left: Space of orbits with boundary defined by 
      $\phi=0$. The vertical plane is a section for constant $J_z$. 
      Middle: Typical section for $N_z=-|\boldsymbol N|,|S_z|<|\boldsymbol S|$.
        This is a part of the continuous family of singular 
        spaces. The singular orbit at the intersection of boundary
        and the constant energy level set has 
             $\mathbb Z_2$ stabilizer. Right: Singular section for 
             $J_z=2|\boldsymbol S|-|\boldsymbol N|$. 
           The singular orbit situated at the intersection of 
          the constant energy level set and the boundary (critical orbit) 
          has stabilizer $SO(2)$.}
    \label{article:fig:sections}
\end{figure}

The idea is to see $H_{\lambda},J_z$ as made up of $SO(2)$-invariant 
polynomials 
\begin{eqnarray*}
    \theta_1&=&S_z\qquad \theta_2=N_z,\qquad \theta_3=N^2_-S_+
           +N^2_+S_-,\\
    \phi&=&N^2_-S_+-N^2_+S_-,
\end{eqnarray*}
satisfying the algebraic relation (syzygy \cite{michel01})
\begin{equation}
\label{article:syzygy}
    \phi^2=\theta_3^2-4(\boldsymbol S^2-\theta_1^2)
             (\boldsymbol N^2-\theta_2^2)^2.
\end{equation}

An orbit of the $SO(2)$  action \eqref{article:groupaction} can be 
characterized by the value of the three algebraically independent invariants 
$\theta_i,i=1,2,3$ and the sign of the linearly independent, but 
algebraically dependent through \eqref{article:syzygy}, invariant 
$\phi$. The space of orbits can then be visualized in a 
$(\theta_1,\theta_2,\theta_3)$-coordinate system as a closed body 
defined by 
\begin{equation}
        \theta_3^2-4(\boldsymbol S^2-\theta_1^2)(\boldsymbol 
                N^2-\theta_2^2)^2\leq 0.
\end{equation}
The space of orbits is shown in \fref{article:fig:sections} 
(left). Its interior points correspond to two orbits distinguished 
by the sign of $\phi$ while the boundary points 
correspond to a single orbit. 

There are three equivalence classes of orbits forming different strata
in the initial $4d$-phase space:
\begin{itemize}
    \item 
    	Generic circular orbits with trivial stabilizer 
            ($4$d regular stratum). 

    \item 
    	A continuous family of orbits for 
        $N_z=\pm |\boldsymbol N|$ and $|S_z|<|\boldsymbol S|$ 
        with stabilized $\mathbb Z_2$ ($2$d critical stratum). 
         These orbits are half as long as a generic orbit. 

    \item 
    	Four isolated critical orbits for 
         $(S_z,N_z)=(\pm |\boldsymbol S|,\pm |\boldsymbol N|)$ 
          with stabilizer $SO(2)$ ($0$d critical stratum). 
\end{itemize}


\subsection{Moment map}

\begin{figure}[h!]
    \begin{center}
    	\includegraphics[width=0.32\textwidth]{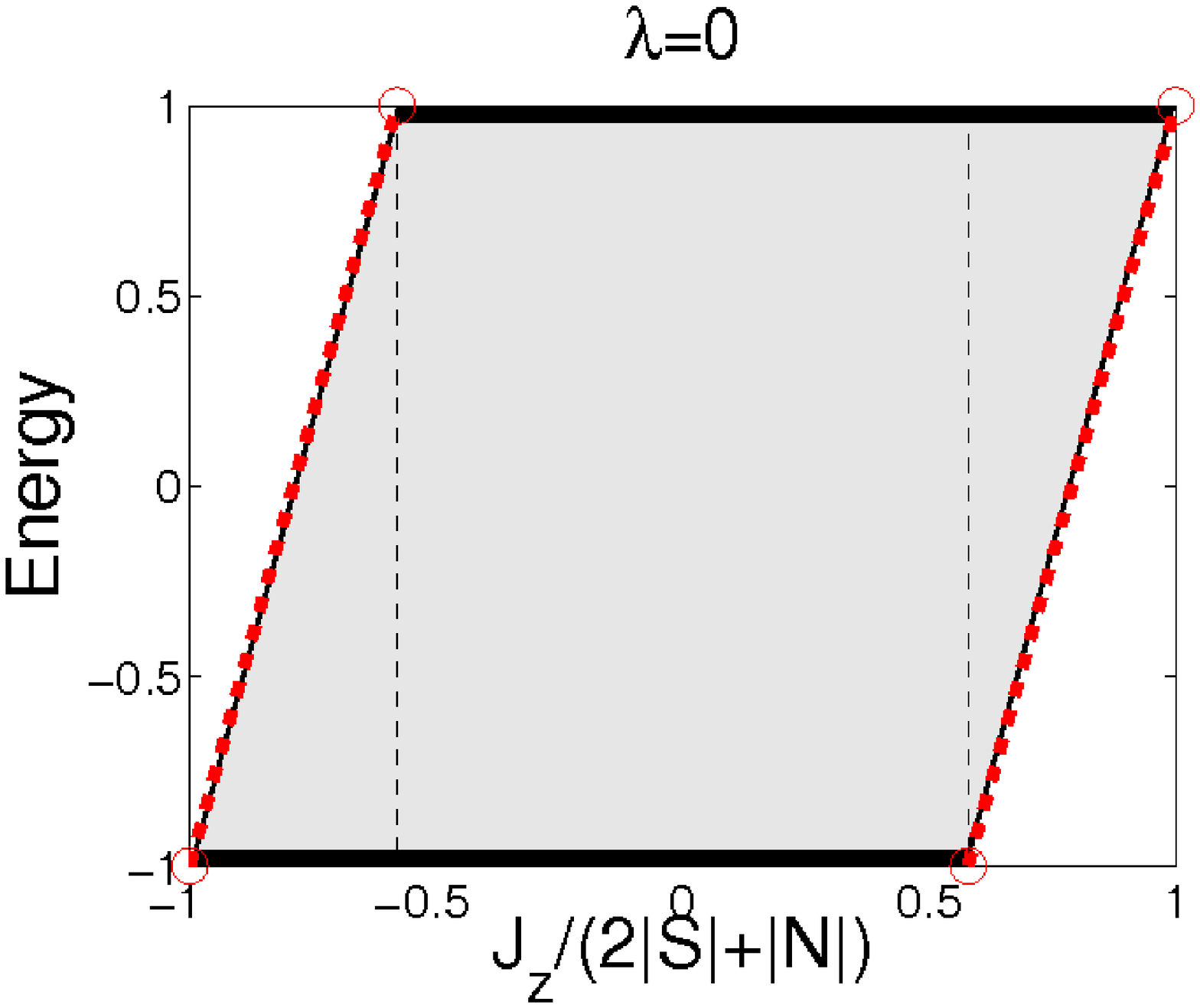}    	    	    	    	 
	\includegraphics[width=0.32\textwidth]{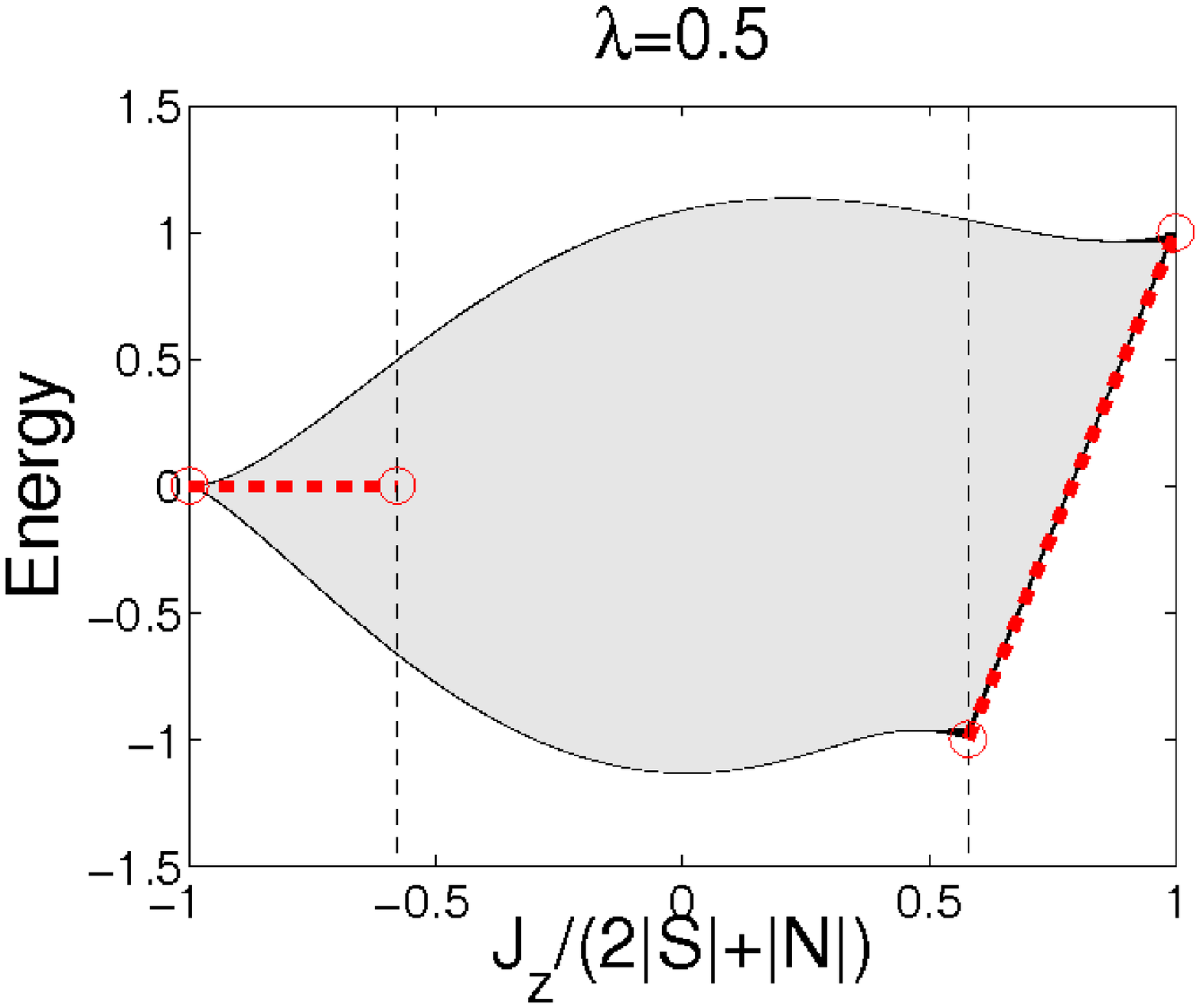}    	
	\includegraphics[width=0.32\textwidth]{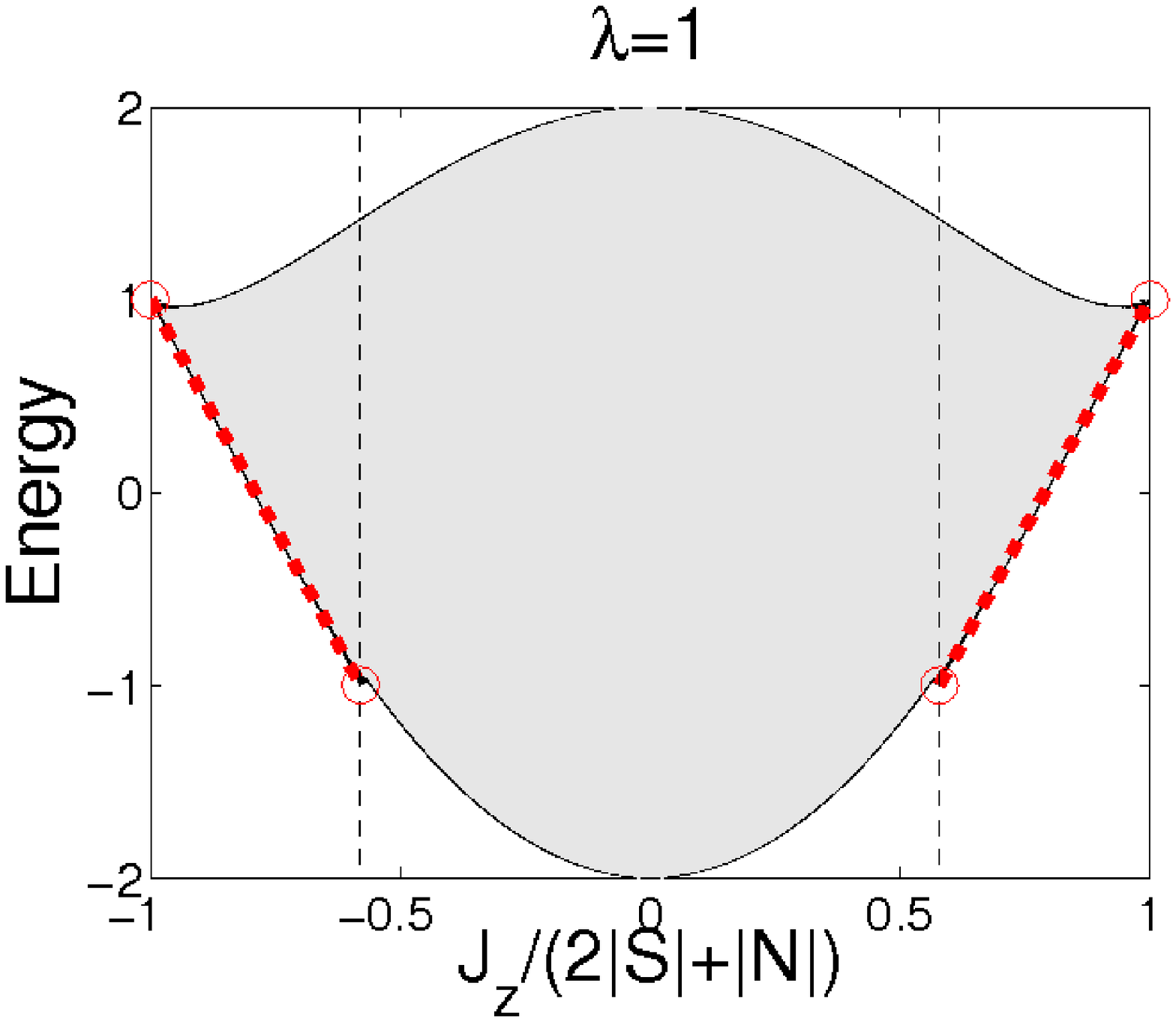}
    \end{center}
    \caption{Image $B_{\lambda}$ of the energy-momentum map 
     \eqref{article:moment} for different values of the external parameter 
      $\lambda$. For $\lambda\sim 1/2$ there are critical values 
       \emph{inside} $B_{1/2}$ and the system has fractional monodromy.}
    \label{article:fig:moment}
\end{figure}

The most natural way to characterize qualitatively 
the classical dynamics for integrable model is 
to introduce the \emph{moment map} \cite{arnold89,guillemin,marsden99}
\begin{equation}
\label{article:moment}
    \boldsymbol F_{\lambda}=(H_{\lambda},J_z):S^2\times 
            S^2\rightarrow\mathbb R^2,
\end{equation}
which maps the compact phase space to a bounded domain 
$B_{\lambda}\subset\mathbb R^2$ which can be expressed as the 
union of regular and critical values of \eqref{article:moment} 
$B_{\lambda}=B^r_{\lambda}\cup B^c_{\lambda}$. Quite naturally 
the shape of $B_{\lambda}$ depends on the parameter $\lambda$ as 
\fref{article:fig:moment} shows.

The moment map defines a fibration over $B_{\lambda}$: for 
fixed $b\in B_{\lambda}$ the dynamics takes place on the fiber 
$\boldsymbol F_{\lambda}^{-1}(h,j)$. Here and later on we use $j$
to denote possible values of $J_z$. 
From the Arnol'd-Liouville theorem it is known that the fiber over a 
regular value $b\in B^r_{\lambda}$ is a $2$-torus \cite{arnold89}. We  
denote it here as a regular fiber. The critical strata in phase space are
mapped via \eqref{article:moment} to the critical values 
$b_c\in B^c_{\lambda}$. These critical values can form isolated points
inside the image of the moment map, boundary lines, or special points
on the boundary, and even lines of critical values situated inside
the image of the moment map. Critical values which belong to the
boundary of the image correspond typically to tori of lower dimension
(circles, or points). Critical values situated inside the image have
nontrivial inverse images  \cite{zung96,zung97,BolFom}.

For $\lambda= \lambda^*$ some of the critical 
values are found in the interior $B_{\lambda^*}$ and such values correspond 
to nontrivial fibers responsible for the appearance of 
\emph{fractional monodromy} \cite{nekoroshev03,nekoroshev03b,KECushSad}.

It is convenient to make a coordinate transformation in the space of orbits
\begin{equation}
	J_z=2S_z+N_z=2\theta_1+\theta_2,\qquad K_z=S_z-2N_z=
                  \theta_1-2\theta_2,
\end{equation}
where $K_z$ is the variable varying on $J_z$-sections.

\Fref{article:fig:sections} shows singular $J_z$-sections together with
constant level sets of energy. It is easy to see geometrically that 
in order to have critical values on the image of the energy-momentum map
inside the domain of regular values it is necessary that the energy 
level going through the singular orbit intersects the boundary of the
orbit space at the singular orbit. In other words we need to compare
the slope of the constant energy level at the singular orbit with
the slope of two boundary lines of the $J_z$-constant section at 
singular point on the boundary. 

It should be noted that at the critical orbit the geometrical form
of the $J_z$ section $\pm (-2|\boldsymbol N|+|\boldsymbol S|-K_z)^{3/2}$
implies that the two boundary lines form the cusp and have the same 
zero slope.
Due to that, the energy section going through critical orbit  intersects
the boundary only if the energy section has itself the zero slope
at critical orbit and this can happen only for $\lambda=1/2$. 
The typical  images of the energy momentum map for $\lambda<1/2$,
$\lambda=1/2$, $\lambda >1/2$ are shown in \fref{article:fig:moment}.
We do not go into details of the evolution of the line of singular values
(dashed red line in \fref{article:fig:moment}) near $\lambda=1/2$
which are related to the possible appearance of second connected
component in the inverse image of the EM map. We note only that 
such complication (as compared with more simple scenario of the
appearance of integer monodromy \cite{sadovskii99}
through Hamiltonian Hopf bifurcation \cite{KEbook}) is due to the presence
of the cusp singularity in the space of orbits. Moreover, it is not
essential for the appearance of the line of singular values together with
the end point inside the EM image as shown in \fref{article:fig:moment},
center, which is responsible for the presence of fractional monodromy. 

 
\subsection{Integer monodromy: Holonomy of the lattice bundle}
 
For each regular value $b\in B^r_{\lambda}$ the periodicity of the 
Arnol'd-Liouville tori defines a $2$d-lattice $\mathcal L_b$ isomorph 
to the regular lattice $\mathbb Z^2$ \cite{arnold89}. Over critical 
values $b_c\in B^c_{\lambda}$ the fiber is singular and we no longer 
have a well-defined lattice. To detect the presence of singular fibers 
it is sufficient to consider the lattice bundle \cite{bates97}
\begin{equation}
    \mathcal L:\bigcup_{b\in \Gamma}\mathcal L_b\rightarrow B^r_{\lambda},
\end{equation}
restricted to a loop $\Gamma:[0,1]\rightarrow B^r_{\lambda}$ in 
$B^r_{\lambda}$. This loop passes only through regular values. 
As $\Gamma(0)=\Gamma(1)$ lifting of $\Gamma$ induces 
an automorphism on fibers, $Aut(\mathcal L_{b=\Gamma(0)})\in 
SL(2,\mathbb Z)$. The bundle $\mathcal L|_{\Gamma}$ depends only on the 
homotopy type of $\Gamma$ such that we only have to consider equivalence 
classes of loops (the fundamental group), $\pi_1(B_{\lambda})$. The 
monodromy map is now defined as 
\begin{equation}
\label{article:monodromy}
    \bmu:\pi_1(B_{\lambda})\rightarrow SL(2,\mathbb Z),
\end{equation}
which is an example of the holonomy concept 
\cite{bates97,nakahara90}.\footnote{Holonomy has become a unifying 
concept in physics, e.g. the Berry phase is seen as the holonomy of a 
$U(1)$-bundle \cite{berry84a,simon83}.} Note that here $\mathcal L$ is 
a flat bundle, i.e. its curvature tensor vanishes.


When the system has an isolated critical value, $\pi_1(B^r_{\lambda})=
\mathbb Z$. The corresponding monodromy map depends on
the topology of the singular fiber and results in the
transformation of  basis cycles of regular tori which can be expressed
as a linear combination with integer coefficients. This gives 
standard integer monodromy  \cite{duistermaat80,bates97,zung97} .

As opposed to almost all previous examples in the literature we no longer 
have isolated critical values. This is shown in 
\fref{article:fig:localMoment} where the critical value
\begin{equation}
    b_c= \boldsymbol F_{\lambda}\Big((0,0,|\boldsymbol N|),
                   (0,0,-|\boldsymbol S|)\Big),
\end{equation}
of the EM map 
is connected to a line, $l_c$, of critical values.
In such a case we have 
$\pi_1(B^r_{\lambda})=0$ for every $\lambda$ as seen from 
\fref{article:fig:moment}, so there is no integer monodromy. However, 
a suitable restriction of the monodromy map \eqref{article:monodromy} 
allows to use closed paths crossing critical line and surrounding critical
value $b_c$. Transformation of the basis cycles of regular tori
after their parallel transfer along such closed paths leads to the new
notion of fractional monodromy
\cite{nekoroshev03,nekoroshev03b,KEbook,KECushSad}.


\subsection{Fractional monodromy: Restriction of basis cycles}
\label{article:sec:fmonodromy}

To determine the fractional monodromy map we have to describe how the 
fibers are \emph{continuously} modified as we go along 
the closed path $\Gamma$ in the base space $B_{\lambda}$ 
of the integrable fibration
and  how the line of critical values can be crossed using only a 
subgroup of cycles generating the fibers. 

\begin{figure}[h!]
\begin{center}
    \input{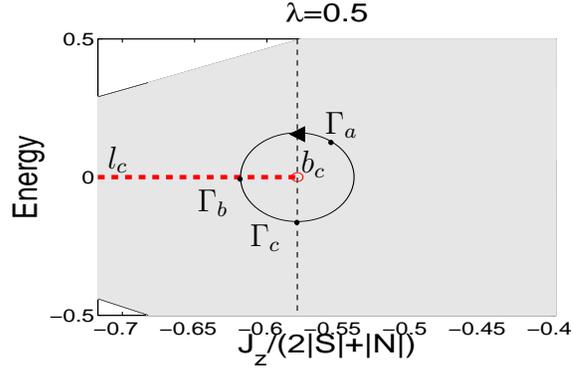}
\end{center}
\caption{The local setup in the image of the moment map $B_{\lambda^*}$ 
when the system has fractional monodromy. The line  $l_c$
of critical values is the projection of  the critical stratum
formed by curled tori \cite{nekoroshev03,nekoroshev03b}. The
critical value $b_c$ is the projection of curled pinched torus which
is the fiber with critical point 
$(0,0,-|\boldsymbol N|),(0,0,|\boldsymbol S|)$.
$\Gamma_a,\Gamma_b,\Gamma_c$ are points on the loop $\Gamma$ associated 
to the fibers represented 
 in \fref{article:fig:crossing}. The figure is done for the ratio
$J/S=15/2$. }
\label{article:fig:localMoment}
\end{figure}

\begin{figure}[h!]
    \begin{center}
    	\subfigure[Fiber over point $\Gamma_a$ 
        (see  \fref{article:fig:localMoment}). 
           The loops are chosen to insure the  continuity of evolution
           along the path $\Gamma$  (see \fref{article:fig:localMoment}),
          especially when crossing singular fiber $\Gamma_b$. ]
                   {\label{article:fig:fatCycles}
                  \includegraphics[width=.4\textwidth]{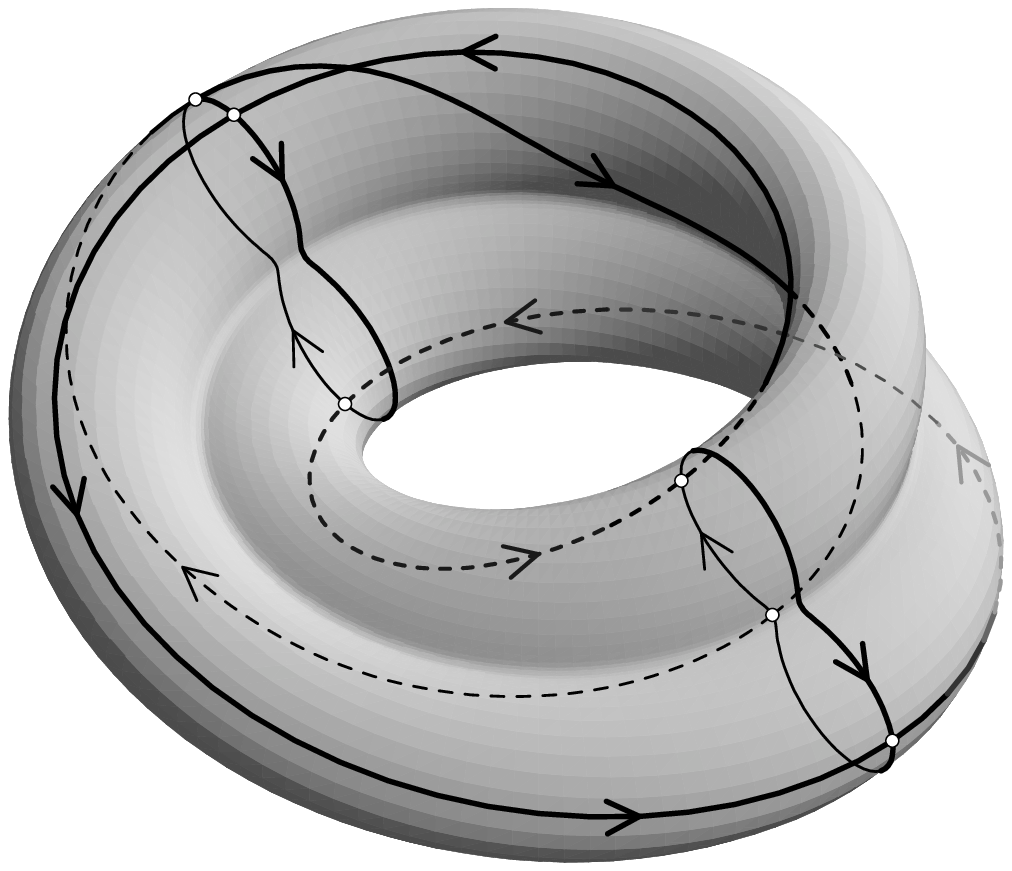}}
    	\subfigure[Fiber over point $\Gamma_b$ (see  
    \fref{article:fig:localMoment}). Intersection of this fiber by an
    auxiliary plane, which is chosen to define the second basic cycle, leads to
     figure eight curve. 
    Generic periodic trajectory of the action intersects twice  
     figure eight.]
               {\label{article:fig:curledCycles}
        \includegraphics[width=.4\textwidth]{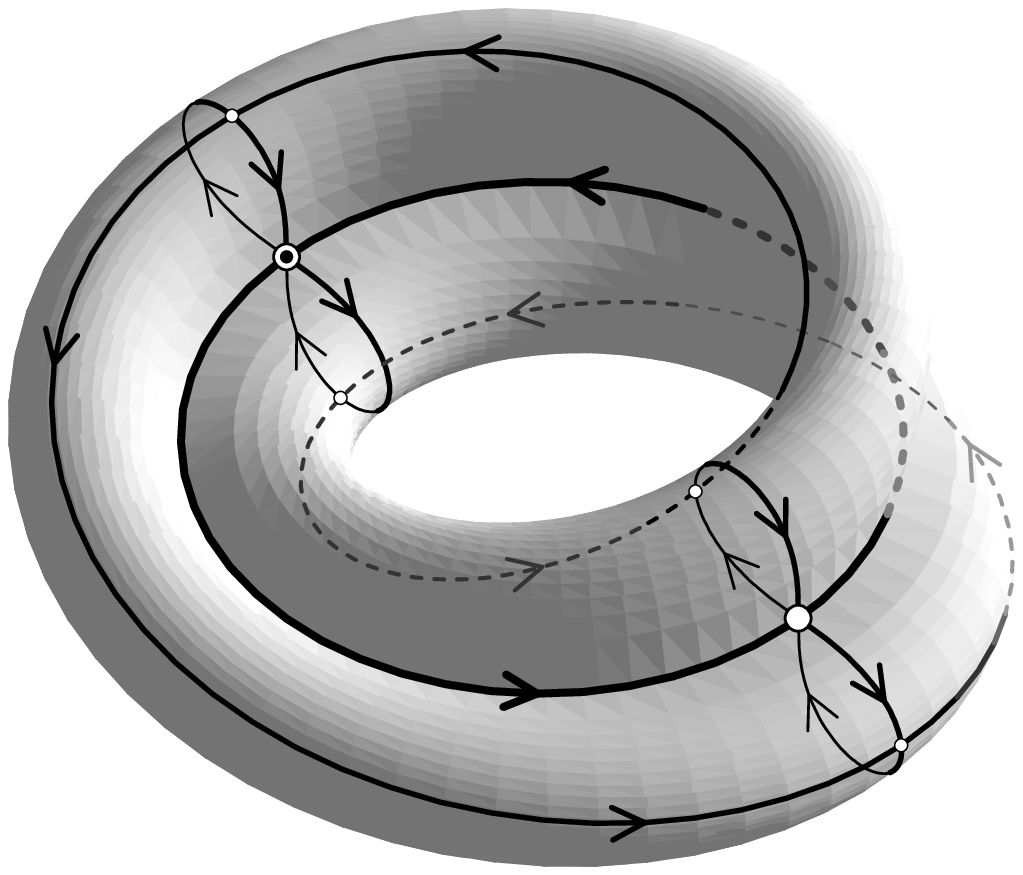}}
    	\subfigure[Fiber over point $\Gamma_c$ 
          (see \fref{article:fig:localMoment}). 
         Loop representing second generating cycle splits into two connected 
         components. This forces to restrict the monodromy map to an index 
              $2$ subgroup of the first homology group of a regular
          fiber which is the origin of fractional monodromy.]
                      {\label{article:fig:slimCycles}
        \includegraphics[width=.4\textwidth]{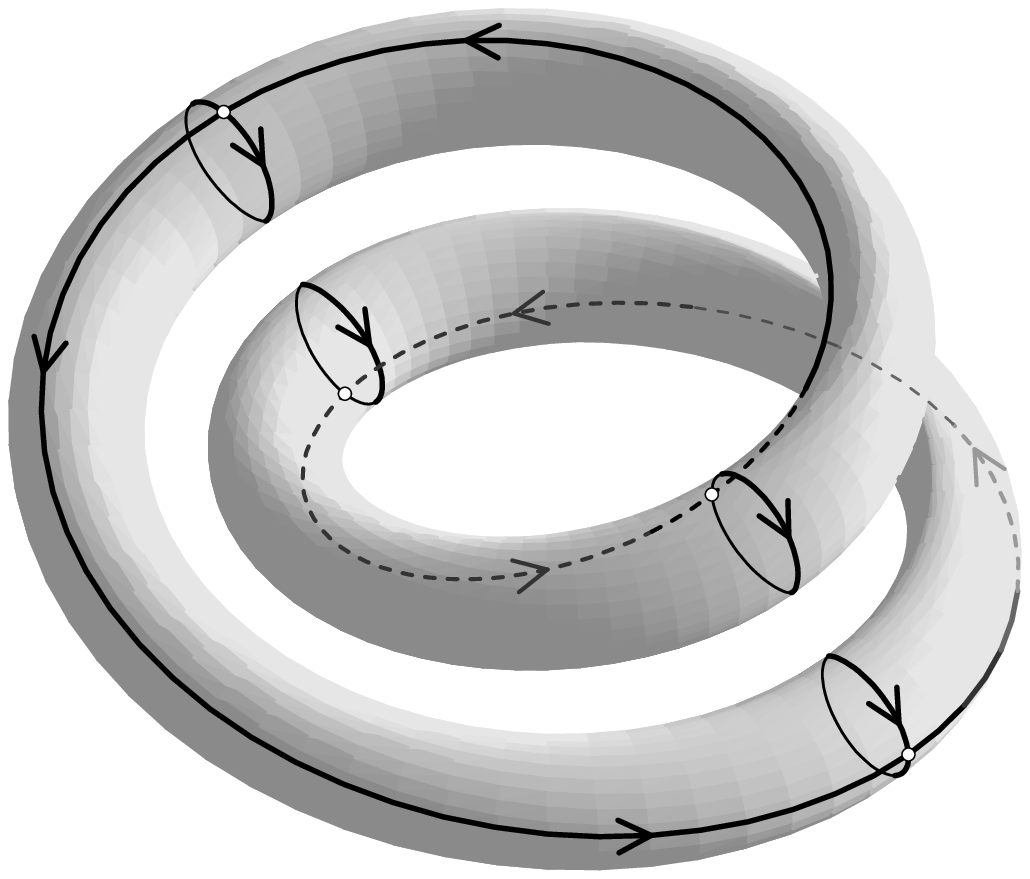}}
    \end{center}
    \caption{Modification of the torus fibers and associated evolution of
loops representing the basis cycles along the path 
$\Gamma$ (\fref{article:fig:localMoment}) as the critical line $l_c$ 
is crossed. Figures taken from \cite{nekoroshev03b}.
 See text for details.}
    \label{article:fig:crossing}
\end{figure}

The local setup in $B_{\lambda}$ is sketched in figures
\ref{article:fig:localMoment} and \ref{article:fig:crossing}.
\Fref{article:fig:crossing} shows the 
fibers at points $\Gamma_a,\Gamma_b$, and $\Gamma_c$ along the loop 
$\Gamma$. In order to understand the evolution of basis cycles
of tori along the contour $\Gamma$  we need to note that  
the trajectories of $J_z$ are closed and well-defined along all  
$\Gamma$. They are due to the $SO(2)$ symmetry of the problem and can 
be used to represent the first of the two cycles generating the 
first homology group of regular fibers. 

The second cycle is chosen as the intersection of  fibers with 
an auxiliary plane. Details of this construction 
are given in \cite{nekoroshev03b}. To pass 
continuously along $\Gamma$ this cycle has to be a double loop as shown in 
\fref{article:fig:crossing}. The main point to notice is the splitting 
of the second generating cycle into two connected components 
(\fref{article:fig:slimCycles}). The applicability of the previous discussion
of fractional monodromy \cite{nekoroshev03b} to the case of the model
Hamiltonian \eqref{article:hamiltonian} studied in the present work 
is confirmed by reducing the model Hamiltonian
$H_{\lambda},J_z$ to the normal form of fractional
monodromy presented in \cite{nekoroshev03b}. This is done in 
\ref{article:app:normalform}.

Due to the splitting of one of the basis cycles when crossing the 
singular stratum, the monodromy map is only defined 
for an index $2$ subgroup of the first homology group of regular fibers. 
This is the essence of 
fractional monodromy. The relation between initial 
basis cycles, $\gamma_{1,2}$, and basis cycles at the end of cyclic evolution, 
$\gamma_{1,2}'$, can be written in the matrix form as \cite{nekoroshev03}
\begin{equation}
\label{article:monodromyNoscaling}
    \left(\begin{array}{c}
		\gamma_1'\\
		2\gamma_2'
	\end{array}\right)=
	\underbrace{
		\left(\begin{array}{cc}
    		1    & 0\\
		-1 & 1  
    	\end{array}\right)}_{\bmu_{\textrm{cl}}}
	\left(\begin{array}{c}
		\gamma_1\\
		2\gamma_2
	\end{array}\right).
\end{equation}
A formal extension of the monodromy map to the basis of the whole 
homology group of regular fibers 
introduces fractional coefficients and a monodromy 
matrix 
\begin{equation}
    \bmu_{\textrm{cl}}=\left(
    \begin{array}{cc}
    	1    & 0\\
	-1/2 & 1  
    \end{array}\right)\in SL(2,\mathbb Q).
\end{equation}

This implies that the preimage $\boldsymbol F^{-1}_{\lambda^*}(\Gamma)$ 
does \emph{not} factorize as $\mathbb T^2\times S^1$ and hence the 
momentum map is \emph{not} a
 principal $\mathbb T^2$-fiber bundle \cite{bates97}. There is then no 
unique way of labeling tori in a vicinity of the pinched curled torus 
and no \emph{global} set of action-angle coordinates can be introduced.

\section{Quantum monodromy}

The Einstein-Brillouin-Kramer (EBK) quantization introduces quantum 
numbers by picking out a set of regular tori \cite{landau65}
\begin{equation}
\label{article:EBK}
    \int_{\gamma_k}p\ud q=2\pi \hbar (n_k+\alpha_k/4),\qquad k=1,2,
\end{equation}
where $\gamma_k$ are basis cycles, generators of the tori, and $\alpha_k$
are Maslov indices. Given this, 
it is no surprise that classical monodromy manifests itself in  quantum 
systems as \emph{quantum monodromy}. The existence of this property 
was first demonstrated on the quantum spherical pendulum \cite{cushman88} 
and later defined as the dual of classical monodromy 
\cite{ngoc99}.\footnote{This is only strictly true in the semi-classical 
  limit. In such a case  the distance between consecutive points in
the spectrum goes to zero and we recover a continuous description.}

\begin{figure}[h!]
    \begin{center}
    	\includegraphics[width=0.32\textwidth]{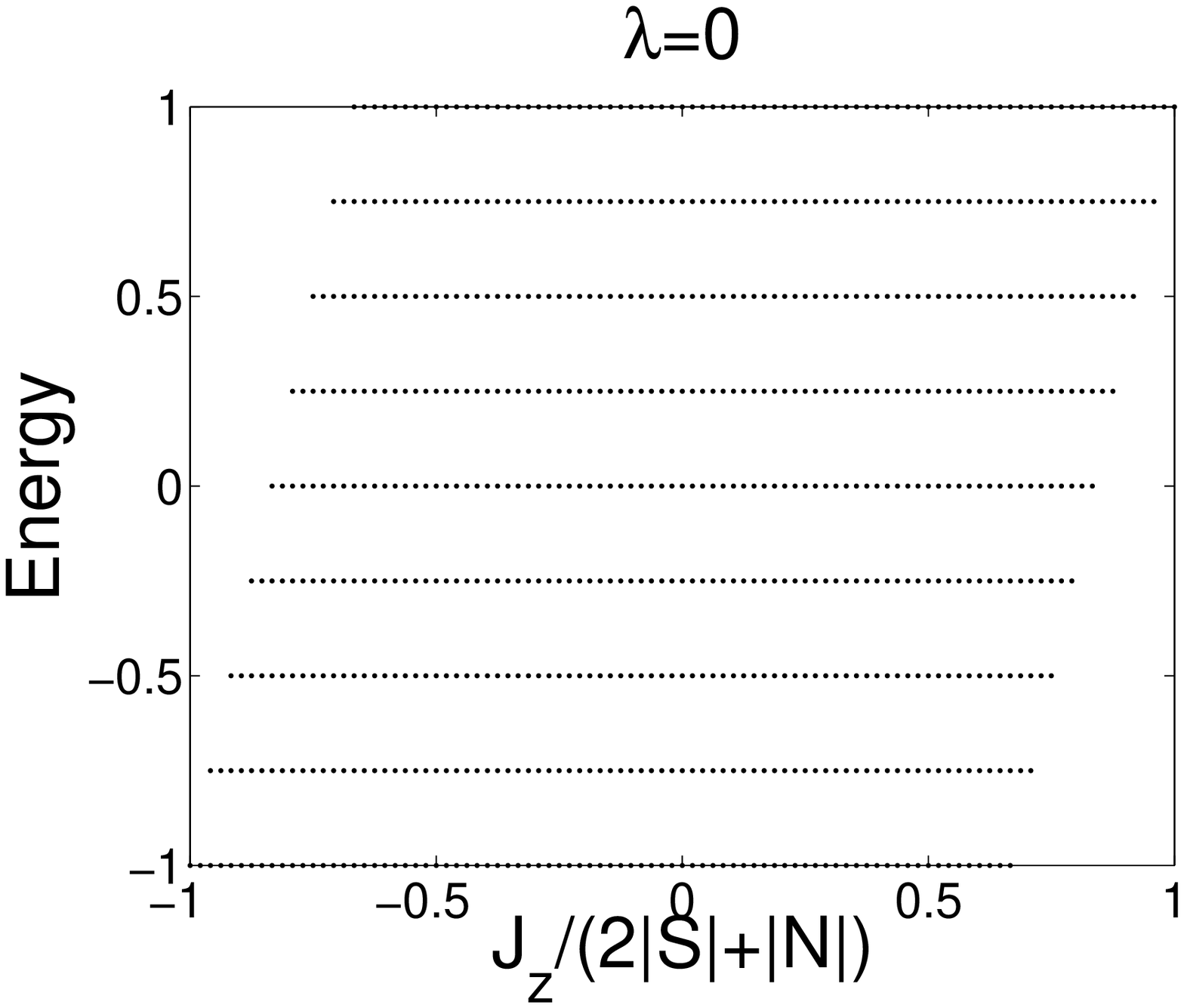}
    	\includegraphics[width=0.32\textwidth]{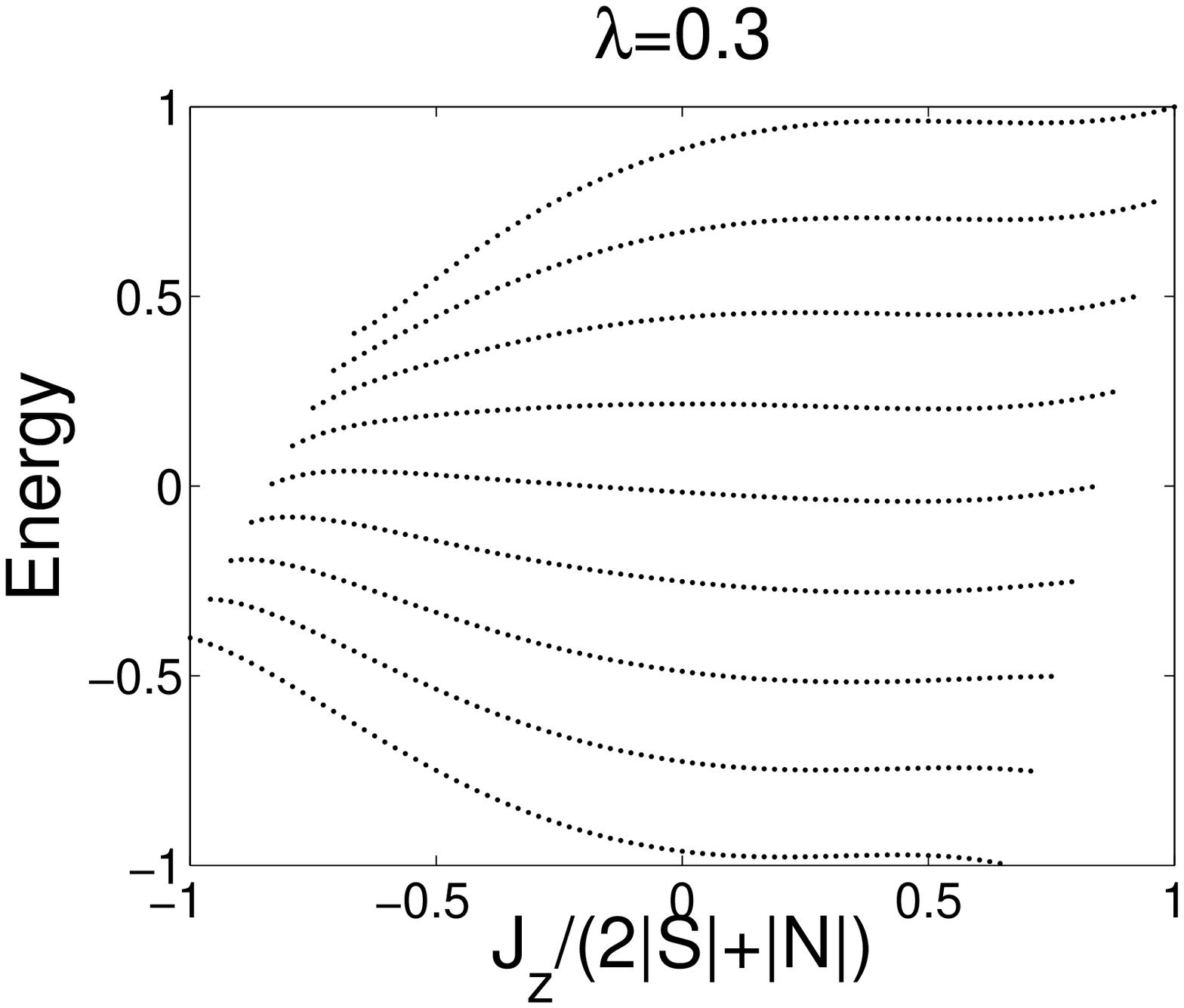}
   	\includegraphics[width=0.32\textwidth]{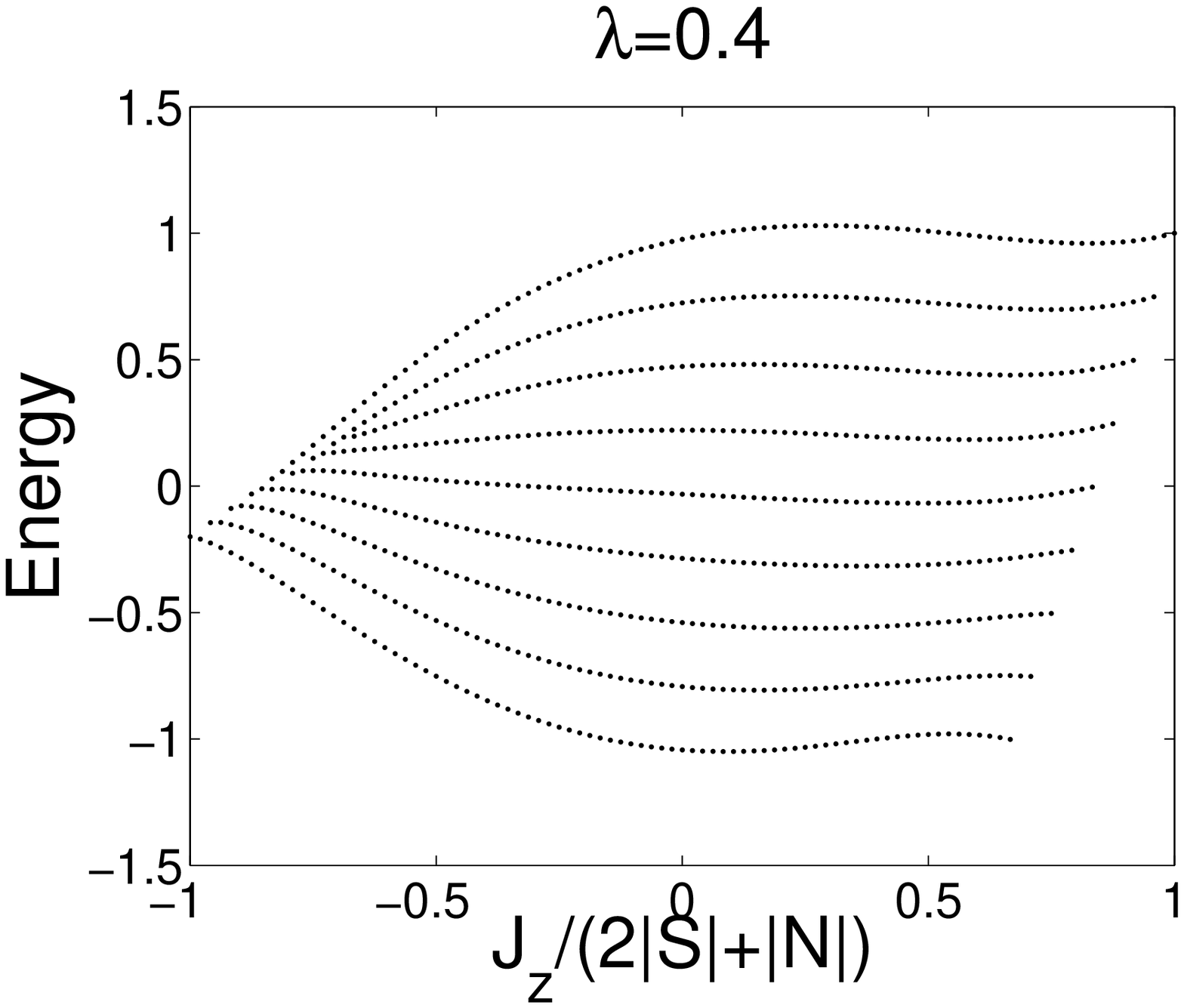}
	
   	\includegraphics[width=0.32\textwidth]{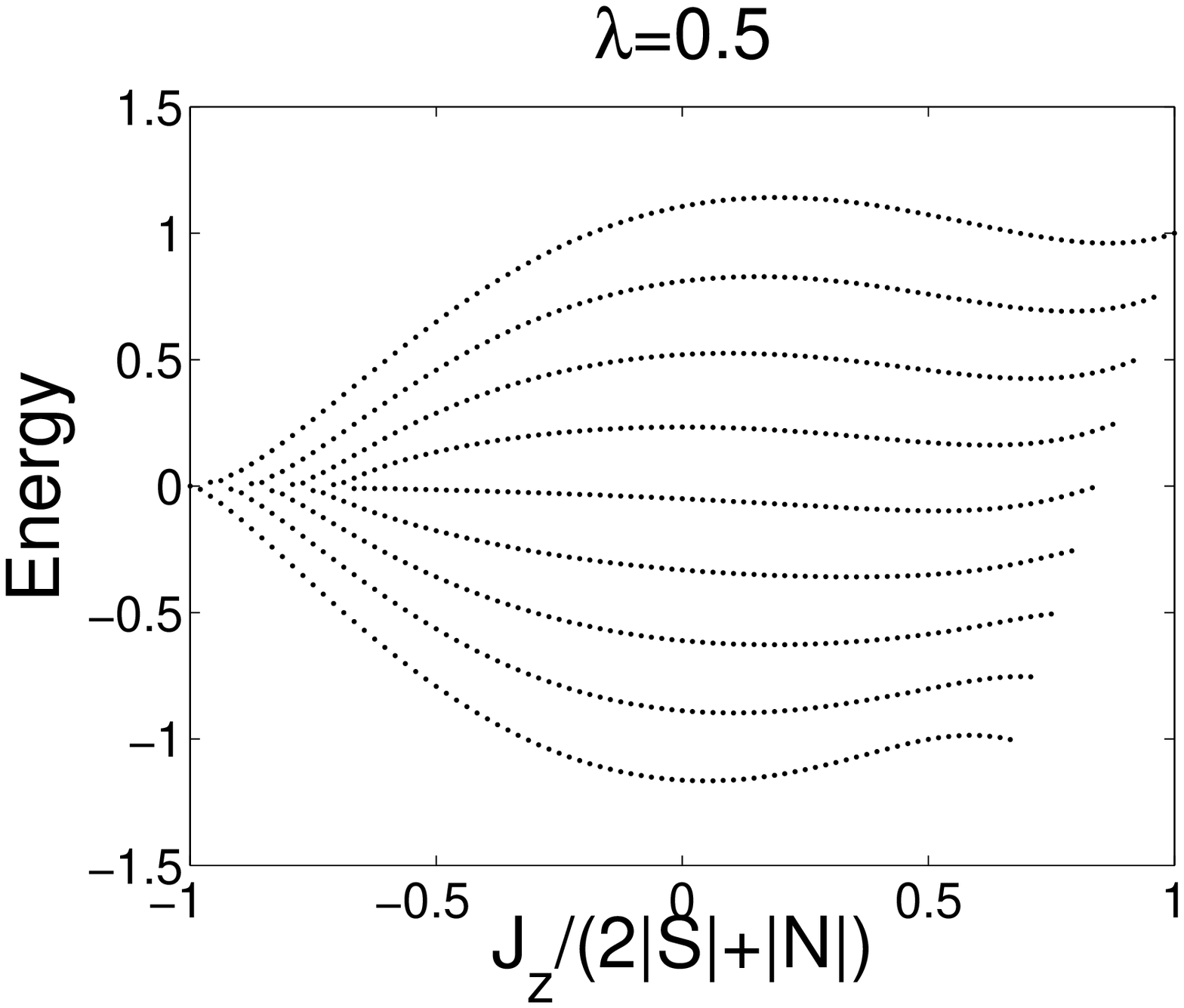}
    	\includegraphics[width=0.32\textwidth]{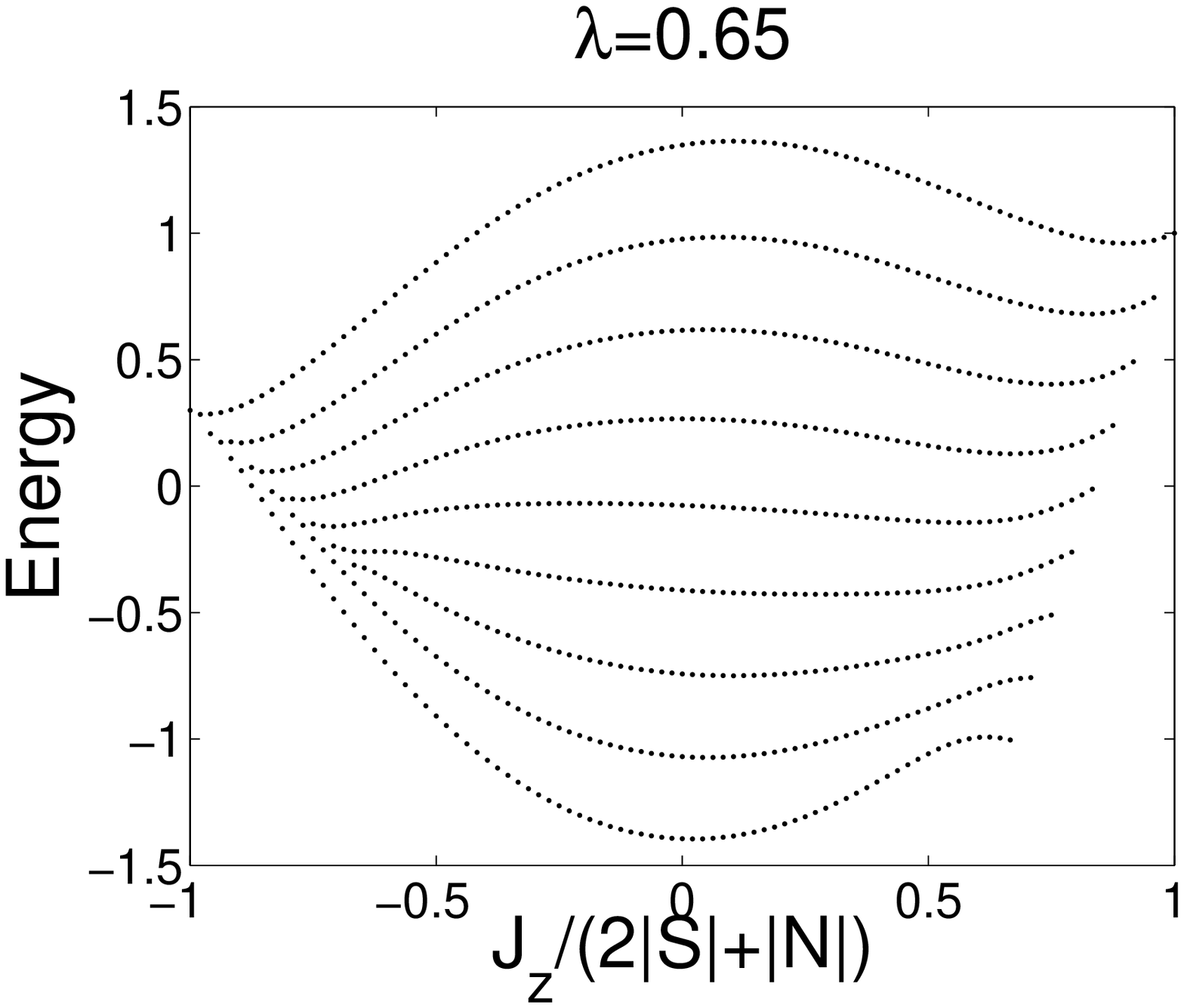}
    	\includegraphics[width=0.32\textwidth]{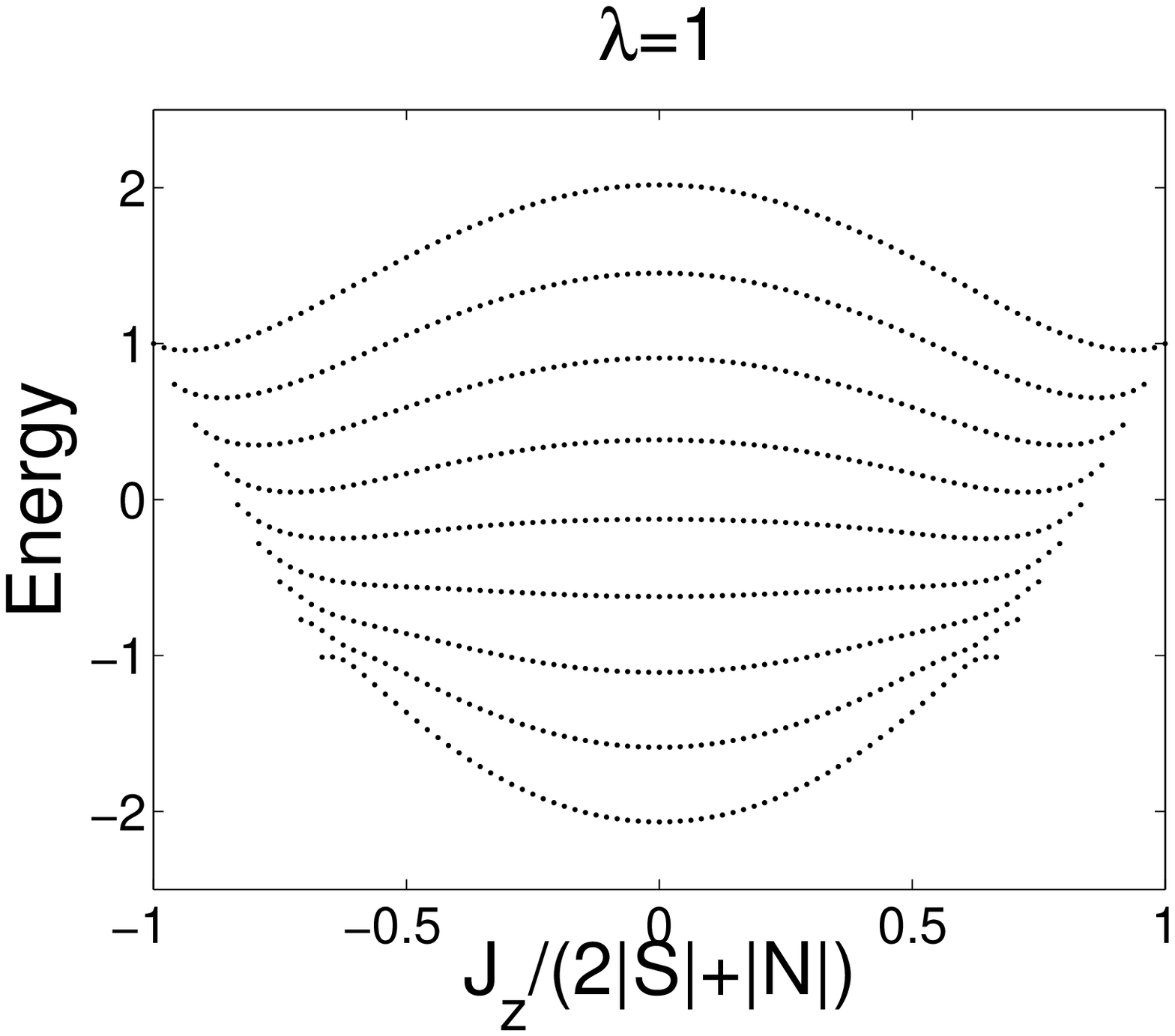}
    \end{center}
    \caption{Modifications of the joint spectrum as $\lambda$ varies,
$\lambda=0\rightarrow 1$. The bands are labeled $\sigma=-S,\dots,S$ from 
the bottom up. For $\lambda=1/2$ there is fractional quantum monodromy
due to the presence of the line of critical values inside the EM map
image. As $1/2 >\lambda\rightarrow \lambda >1/2$ there is a modification of 
the band structure 
due to the displacement of the line of critical values from the
boundary of the EM image into inside and further to the 
new position at the boundary (see \protect{\fref{article:fig:moment}}).
} 
    \label{article:fig:js}
\end{figure}

The EBK rules lead to a $2$d-lattice of quantum states - or joint 
spectrum - in $B_{\lambda}$. 
From \eqref{article:EBK} the distance between consecutive quantum 
states decreases as $\hbar\rightarrow 0$. Our model is a coupling 
of two angular momenta $\boldsymbol S$ and $\boldsymbol N$ with 
effective Planck constants $\hbar_S,\hbar_N$ respectively. The 
assumption $S\ll N$ leads to $\hbar_N\ll \hbar_S$ and to the existence of 
two scales in the joint spectrum. This explains the local band 
structure easily observed in \fref{article:fig:js}. We label the bands 
by the quantum number of $S_z$, $\sigma=-S,\dots,S$. 

For $\lambda=0$, the joint spectrum forms globally a regular lattice which
possesses a well defined (up to a similarity
transformation with $SL(2,Z)$ matrix) elementary cell over the whole lattice. 
This means that there exists  a global 
labeling of states. The lattice remains to be regular
(just in slightly deformed form) for the $\lambda$-dependent family of 
integrable systems up 
to $\lambda\sim 1/2$. At $\lambda=1/2$ the presence of
one-dimensional defect is clearly seen within the regular part of the
lattice. This defect results in a 
modification of the bands. For $\lambda=1$ we again have a globally 
regular lattice but now with a different band structure.

\begin{figure}[h!]
    \begin{center}
        \input{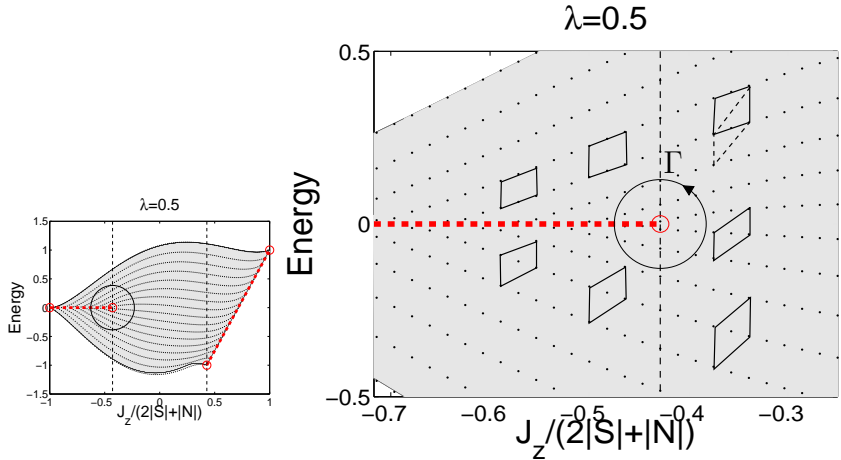}
    \end{center}
    \caption{Joint spectrum for $S=8,N=40$ and $\lambda=1/2$ which
         shows the effect of fractional monodromy.
Left: The global view of the joint spectrum. Right: Parallel
transport of the double cell along a closed path crossing once the
line of critical values and surrounding the critical value
$(J_z=2S-N, E=0)$ of the EM map. For  $S=8,N=40$ we have
 $J_z/(2|S|+|N|) = - 3/7\approx -0.4286$.
} 
    \label{article:fig:jsCritical}
\end{figure}

For $\lambda\sim 1/2$, the effect of the defect on the lattice is 
characterized (up to conjugation) by an element $\bmu_{\textrm{qm}}$ 
determined in the following way (see \fref{article:fig:jsCritical}):

\begin{itemize}
\item
    	Make a choice of cell. To pass the line defect the 
cell should be doubled in $J_z$ direction. This is the quantum 
analogue of the restriction imposed on the choice of passable cycles 
in section 
\ref{article:sec:fmonodromy}. Cell doubling is not necessary
in the case of integer monodromy 
\cite{sadovskii99}.

\item
    Moving along the path $\Gamma$ between initial and final points 
the elementary cell does not change as long as the path remains within the class of homotopically trivial paths.
	 However, after translation along a path $\Gamma$ as shown in \fref{article:fig:jsCritical} we return with a different cell. A rescaling as done in section  \ref{article:sec:fmonodromy} gives
\begin{equation}
\label{article:qmonodromy}
    \boldsymbol{\mu}_{\textrm{qm}}=\left(
    \begin{array}{cc}
    	1 & 1/2 \\
    	0 & 1 
    \end{array}\right)\in SL(2,\mathbb Q),
\end{equation}
which is the quantum monodromy matrix (after a formal rescaling of 
cell).\footnote{Here we observe the duality between classical and 
quantum monodromy explicitly as 
$\bmu_{\textrm{qm}}={^t}(\bmu_{\textrm{cl}})^{-1}$ \cite{ngoc99}.}

\end{itemize}
The non-triviality of monodromy shows  that no unique set of 
quantum numbers exists which can be
used to label states in the joint spectrum \cite{giacobbe04}. 
This is of special importance for molecular 
physics where effective quantum numbers are typically introduced 
on the basis of experimental spectral information using extrapolation
within effective models.

\subsection{Decomposition into sublattices}

\begin{figure}[h!]
    \begin{center}
        \input{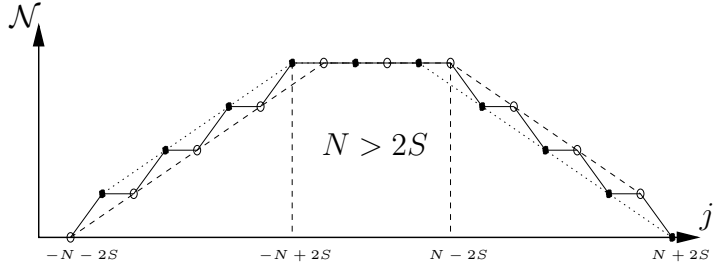}
    \end{center}
    \caption{The number-of-states function $\mathcal N_j$  is a quasipolynomial 
(full line). The existence of two length scales in the system is due to 
the non-diagonal $SO(2)$-action. Retaining only the linear term, i.e. 
restricting to either even ($\circ$) or odd ($\bullet$) values of $j$, 
results in two subsystems with integer monodromy.} 
    \label{article:fig:numberofstates}
\end{figure}

Let $j$ label the eigenvalues of $J_z$, the second integral of motion, 
and $\mathcal N_j$ be the dimension of the associated eigenspace, i.e. 
the number of states with $J_z=const$  on \fref{article:fig:js}. The 
number of states function (\fref{article:fig:numberofstates}) is a 
quasipolynomial, i.e. polynomial in $j$ with coefficients being
periodic in $j$:
\begin{equation}
\label{article:numstates}
	\mathcal N_j=\left\{
	\begin{array}{cc}
		2S+1,                                     & |j|\leq N-2S\\
		\frac 12(J-|j|+\frac 12(3+(-1)^{J+|j|})), & {\rm otherwise}
	\end{array}\right.  .
\end{equation}
This reflects the existence of two different scales in the system. A large 
scale behavior 
is associated with polynomial part, whereas a small scale behavior 
is characterized 
by the oscillating term. This is a direct consequence of the 
non-diagonal $SO(2)$ 
action  as described in section \ref{article:symmetry}.

\begin{figure}[h!]
    \begin{center}
        \input{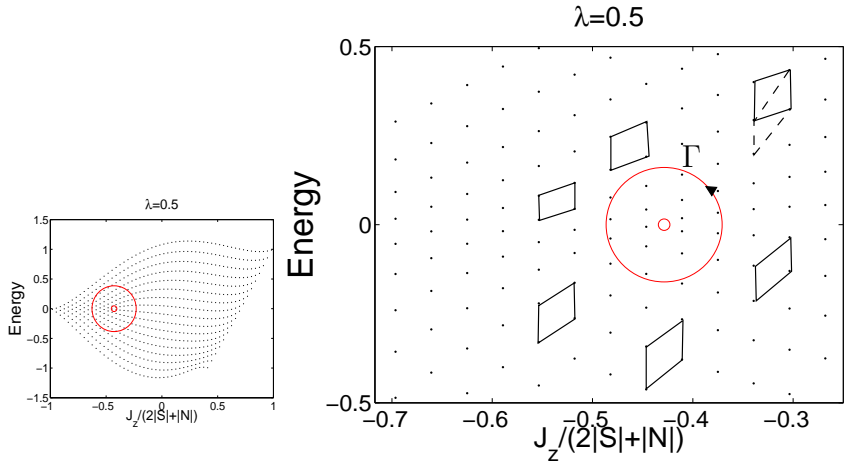}
    \end{center}
    \caption{Index $2$ sublattice of the joint spectrum for odd values 
of $J_z$. This sublattice possesses only one isolated critical
value. The path encircling this critical value is
characterized by integer monodromy.  The situation is similar 
for even $J_z$.} 
    \label{article:fig:subLattice}
\end{figure}

Restricting ourselves to only even or odd values of $J_z$ amounts to 
ignoring the oscillating part of \eqref{article:numstates}. This gives 
integer monodromy on each index $2$ sublattice of the joint spectrum as 
shown in \fref{article:fig:subLattice}. 
Disregarding the small scale 
behavior our system reduces to two systems with $1:(-1)$ resonance of 
the type found in \cite{sadovskii99}.

Integer monodromy on index $2$ sublattices should be compared with the 
monodromy matrix \eqref{article:monodromyNoscaling}, i.e. \emph{before} 
the formal rescaling of the restricted basis cycle. This is another way 
of showing how fractional monodromy can be seen as integer monodromy for 
an appropriate subset of basis cycles.


\subsection{Quantum monodromy and redistribution of states}
\label{article:sec:qmRedistribution}

Returning to \fref{article:fig:js} we observe that the breaking of band 
structure is related to the appearance of monodromy and to a rearrangement 
of bands seen as a transfer of states from the lower to the upper bands. 
Counting the number of states before and after modification of the
position of the singular stratum on the image of EM map gives:
\begin{equation}
    	\Delta\mathcal N_\sigma=4\sigma,\sigma=-S,\dots,S,
\end{equation}
where $\mathcal N_\sigma$ is the number of states in the 
$\sigma$th band (labeled  from the bottom up).

The classical equivalent of the redistribution of the number
of states in bands is a transfer of phase space volume to
higher energies.
This quantum-classical correspondence is 
explained by the EBK rules \eqref{article:EBK} relating the volume of 
the reduced classical phase space to the number of quantum states with 
a given value, $n_k$, of the integral of motion. In both the classical 
and quantum mechanical case monodromy is thus related to a redistribution 
event.


\section{Semi-quantum description: Chern index}

We now proceed to consider the  semi-quantum or Born-Oppenheimer 
description which is valid in the limit $S<<N$. Here the slow motion 
of $\boldsymbol N$ is classical, and for any given value of 
$\boldsymbol N$ the fast motion of $\boldsymbol S$ is quantum mechanical 
and dependent on $\boldsymbol N$. The fast motion is generated by the 
Hamiltonian $\hat H_{\boldsymbol N,\lambda}$ acting in $\mathcal H_S$ 
and obtained by substituting the operators $\hat{\boldsymbol N}$ by 
the classical variable $\boldsymbol N\in S^2$. 

This operator has normalized eigenstates 
\begin{equation}
    \hat H_{\lambda,\boldsymbol N} \ket{\psi_\sigma(\lambda,\boldsymbol N)}=
      E_\sigma(\lambda,\boldsymbol N)\ket{\psi_\sigma(\lambda,\boldsymbol N)}.
\end{equation}
with $\sigma=-S...+S$. The eigenvalues, 
$E_\sigma(\lambda,\boldsymbol N):S^2\rightarrow\mathbb R$ , 
seen as functions of $\boldsymbol N$ form $2S+1$ bands calculated in 
the following way:

The quantum Hamiltonian is an operator valued symbol 
\begin{eqnarray}
\label{article:semiquantumSymbol}
    \boldsymbol N\in S^2&\mapsto&\hat H_{\lambda,\boldsymbol S}= 
      \boldsymbol K_{\lambda}(\boldsymbol N)\cdot 
            \frac{\hat{\boldsymbol S}}{|\boldsymbol S|},\\
    \boldsymbol K_{\lambda}(\boldsymbol N)&=&\left(
         \frac{2\lambda}{|\boldsymbol N|^2}\left(N_+^2+N_-^2\right),
           \frac{-2i\lambda}{|\boldsymbol N|^2}\left(N_+^2-N_-^2\right),
      (1-\lambda)+\frac{\lambda}{|\boldsymbol N|}N_z \right),\nonumber
\end{eqnarray}
defined by
\begin{equation}
    \bra{\boldsymbol N}\hat H_{\lambda}\ket{\boldsymbol N}=
\hat H_{\lambda,\boldsymbol N} + O(\epsilon_N),
\end{equation}
the principal symbol with respect to $\boldsymbol N$.

Explicit eigenvectors are constructed from the angular momentum basis 
vectors by applying the rotation taking the $z$-axis into 
$\boldsymbol K_{\lambda}$
\begin{equation}
\label{article:eigenvectors}
    \ket{\psi_\sigma(\lambda,\boldsymbol N)}=
       e^{\boldsymbol K_{\lambda}(\boldsymbol N)\cdot 
           \hat{\boldsymbol S}}\ket{\sigma},\sigma=-S,\dots,S,
\end{equation}
and the spectrum is 
\begin{eqnarray}
\label{article:bandHamiltonian}
    E_\sigma(\lambda,\boldsymbol N)&=&
            \frac{\sigma}{|\boldsymbol S|}|\boldsymbol 
                   K_{\lambda}(\boldsymbol N)| \nonumber \\
    \label{nfast:energySurface}	
    &=& \frac{\sigma}{|\boldsymbol S|}
          \sqrt{\left(\frac{4\lambda}{|\boldsymbol N|^2}\right)^2
                \left(N_x^2+N_y^2\right)^2+\left((1-\lambda)+
                    \frac{\lambda}{|\boldsymbol N|}N_z\right)^2},
\end{eqnarray}

which shows us that the only degeneracy between bands occurs for 
\begin{eqnarray}
    N_x=N_y=0\Rightarrow N_z=\pm |\boldsymbol N|,\\
    (1-\lambda)+\lambda \frac{N_z}{|\boldsymbol N|}=0,
\end{eqnarray}
with only solution $(\lambda^*,\boldsymbol N^*)=
(1/2,(0,0,-|\boldsymbol N|))$. In this case there is a  collective 
degeneracy between \emph{all} bands in the semi-quantum spectrum due 
to the high degree of symmetry of the model 
\cite{pavlov88,sadovskii99}.\footnote{$k$th order eigenvalue 
degeneracies of a Hermitian operator occur in a space of dimension 
(dim$_{parameters}-(k^2-1)$) \cite{avron88}. With three independent 
parameters $(\lambda,\boldsymbol N)\in [0,1]\times S^2$ only point-wise 
degeneracies between pairs of eigenvalues are generic, i.e. cannot be 
removed by perturbing the model. The important point is that with 
three parameters
we shall always have band degeneracies where Chern index can be 
"exchanged" \cite{faure01}.}


\subsection{Complex line bundles over $S^2$}

For each $\sigma=-S,\dots,S$ there is a natural vector bundle structure 
associated to a parameter dependent operator constructed as follows:

The normalized eigenvectors \eqref{article:eigenvectors} are only 
defined up to a phase factor but the projector
\begin{equation}
    \hat P_\sigma:\boldsymbol N\in S^2\mapsto 
   \ket{\psi_\sigma(\lambda,\boldsymbol N)}\bra{\psi_\sigma(\lambda,\boldsymbol N)},
\end{equation}
onto the corresponding eigenspace is well-defined and associates to 
each point $\boldsymbol N\in S^2$ a one dimensional complex subspace 
of $\mathcal H_S$. This defines $2S+1$ complex line bundles 
$L_s\rightarrow S^2$ for almost all values of $\lambda$ (except when 
the degeneracy mentioned in the previous section is encountered). Each bundle 
has an isomorphism class depending on $\lambda\in [0,1]$ and 
characterized by a single integer $C_\sigma\in\mathbb Z$, the so-called 
Chern index  \cite{griffiths78,faure01}.


\subsection{Trivial topology}

For $\lambda=0$ eigenstates form the usual angular momentum basis 
set $\ket{\psi_\sigma(0,\boldsymbol N)}=\ket{\sigma}$. As these states are parameter 
independent we have $2S+1$ trivial line bundles over $S^2$ characterized 
by $C_\sigma=0$. As the topology remains unchanged under continuous 
deformations this remains true until the sphere spanned by $\boldsymbol N$ 
encounters $(\lambda^*,\boldsymbol N^*)$ at the south pole.

This happens for $\lambda=1/2$ and the collective degeneracy can be seen 
as a trivial rank$_{\mathbb C}$ $2S+1$ bundle over $S^2$. In fact, since 
the total space $\mathcal H_S$ is a trivial vector bundle 
\begin{equation}
    C=\sum_{\sigma=-S}^S C_\sigma=0.
\end{equation}
for all values of $\lambda$.


\subsection{Nontrivial topology}

As the only degeneracy occurs at $(\lambda^*,\boldsymbol N^*)$ it is 
sufficient to calculate $C_\sigma'$ for $\lambda=1$. This is done algebraically 
by defining the Chern index $C_n'$ as a sum of oriented zeroes of a 
global section \cite{faure01}.\footnote{A section is a continuous choice 
of element in each fiber. A non-vanishing section globally defines a 
frame and hence a global separation of the bundle. In this case 
$S^2\times \mathbb C$ and the bundle is said to be trivial 
\cite{griffiths78}.}

A choice of a reference coherent state $\ket{\boldsymbol S_0}$ defines 
a global choice section
\begin{equation}
    \hat P_\sigma(\boldsymbol N)\ket{\boldsymbol N_0}=
          \ket{\psi_\sigma(1,\boldsymbol N)}\bra{\psi_\sigma(1,\boldsymbol N)}
              \boldsymbol S_0\rangle,
\end{equation}
where $\hat P_\sigma(\boldsymbol N)$ is the projector onto the $\sigma$-th 
eigenspace in $\mathcal H_S$ spanned by $\ket{\psi_\sigma(1,\boldsymbol N)}$. 
The section has the same zeroes as the Husimi distribution
\begin{equation}
\label{article:husimi}
    \mathcal H_\sigma(\boldsymbol S)=|\bra{\psi_\sigma(1,\boldsymbol N)}
                     \boldsymbol N_0\rangle|^2,
\end{equation}
of $\ket{\boldsymbol S_0}$. Here $\ket{\psi_\sigma}$ is simply a rotation of 
the angular momentum eigenstates $\ket{\sigma}$ with a Husimi distribution 
known to have $(S-\sigma)$ oriented zeroes at $\boldsymbol K_1(\boldsymbol N)$ 
and $-(S+\sigma)$ oriented zeroes at $-\boldsymbol K_1(\boldsymbol N)$ 
\cite{leboeuf91}.

Introducing spherical coordinates $(\Phi,\Theta)$ on parameter sphere
\begin{equation}
    \boldsymbol K_{1}(\Phi,\Theta)= \left(4\sin^2(\Theta)\cos(2\Phi),
            4\sin^2(\Theta)\sin(2\Phi),\cos(\Theta)\right),
\end{equation}
we see that as $(\Phi,\Theta)$ cover the sphere once $\ket{\psi_\sigma}$ 
cover phase space twice. Then each set of zeroes pass over all points 
on the sphere - including $\boldsymbol S_0$ - twice and 
\begin{equation}
    C_\sigma'=2\left(S-\sigma+(-(S+\sigma))\right)=-4\sigma.
\end{equation}
The change in Chern index for the $\sigma$-th bundle is then 
\begin{equation}
\label{article:dChern}
    \Delta C_\sigma=C_\sigma'-C_\sigma=-4\sigma,
\end{equation}
as $\lambda=0\rightarrow 1$. 


\subsection{Exchange of states and indices: An index formula}

In section \ref{article:sec:qmRedistribution} the change in number of 
states was found to be $\Delta\mathcal N_\sigma=4\sigma$ such that
\begin{equation}
    \Delta C_\sigma=-\Delta \mathcal N,
\end{equation}
and $\mathcal N_\sigma+C_\sigma$ is conserved for all values of $\lambda$. 
When $\lambda=0$ we have $\hat H_0(\boldsymbol N)=\hat S_z$ and 
$\mathcal N_\sigma=2S+1=$dim$\mathcal H_N$ which leads to
\begin{equation}
\label{article:indexFormula}
    \mathcal N_\sigma=\textrm{dim}\mathcal H_S-C_\sigma,
\end{equation}
relating the topology of a complex line bundle in the semi-quantum 
description to the number of quantum states in a band  \cite{faure00}. 
This so-called index formula on the sphere is the simplest case of the 
Atiyah-Singer index formula \cite{fedosov00}. 
\section{Discussion}

Quantum systems with a slow-fast coupled motion are very common in nature, 
the textbook example being that of a rovibrational molecular system 
\cite{pavlov88,sadovskii99,faure00,cushman00,joyeux03,zhilinskii01}. 
We have given a model example of such a system with a specific 
(nondiagonal) action of the dynamical symmetry group which has the 
additional property of being integrable.

The \emph{raison d'\^etre} of our model is an $SO(2)$ with a non-diagonal 
action leading to fractional monodromy, the essence being a restriction 
of the monodromy map to an index $2$ subset of basis cycles. To our 
knowledge this is currently the only example of fractional monodromy 
in a system with compact phase space. This gives a bounded spectrum which 
is important when we turn to the physically relevant question of 
redistribution.  Hydrogen atom in the presence of electric and magnetic
fields leads under certain conditions to effective models which 
manifest the fractional monodromy effect \cite{ProcRS}.

Here we observe that the appearance of monodromy is related to a breaking 
of the band structure in the joint spectrum. Furthermore this is 
associated to a rearrangement of bands seen as a redistribution of quantum 
states. From the orbit space analysis we see that it makes sense to talk 
about monodromy in the limit of adiabatic coupling 
$|\boldsymbol S|/|\boldsymbol N|\rightarrow 0$. This is yet another fact 
establishing the connection between redistribution and monodromy.

In the semi-quantum description the notion of integrability is not present 
but the redistribution of states appears as a change in the Chern index 
of the associated complex line bundles. This is the result of a simple 
index formula expressing the redistribution of levels in terms of Chern 
indices \cite{faure00}. 

From the semi-quantum analysis we know that redistribution is stable 
under perturbation. Given our hypothesis concerning its relation to 
monodromy it is tempting to assume that the quantum/classical analysis 
can be extended to quasi integrable (KAM) systems. This general extension 
has already been done in the case of integer monodromy 
\cite{broer02,rink04}. For the fractional monodromy though, the 
critical point responsible for the monodromy 
is no longer isolated but connected to a line of hyperbolic points, and this 
makes more difficult the extension to the KAM regime.

Also a recent generalization of the so-called moment polytopes of 
Atiyah, Guillemin-Sternberg and Delzant to problems with integer monodromy 
\cite{ngoc03} makes the precise relation between redistribution 
(Chern index) and general $p/q$-monodromy a pertinent question. Our 
model can easily be generalized to $1/k$-monodromy \cite{hansen04} but 
for the time being the more actual question is to find a physical 
example of system exhibiting fractional monodromy and the 
redistribution phenomenon.

\ack
M.S.H. would like to thank LPMMC exquisite hospitality during his 
masters thesis $(2003/04)$. This work was partly supported by the EU 
project Mechanics and Symmetry in Europe (MASIE), Contract No. 
HPRN-CT-2000-00113.

\begin{appendix}


\section{Local structure of the moment map: Monodromy}
\label{article:app:normalform}

To establish the presence of fractional monodromy $H_{1/2},J_z$ is 
reduced to a normal form for fractional monodromy presented in 
\cite{nekoroshev03}. This is done by linearizing around 
$(\boldsymbol N^*,\boldsymbol S^*)=((0,0,|\boldsymbol N|),
(0,0,-|\boldsymbol S|))$
\begin{eqnarray*}
    N_x=p_1,N_y=q_1,N_z=
    \sqrt{1-(N_x^2+N_y^2)}\simeq 1-\frac12 \left(p^2_1+q^2_1\right),\\
    S_x=p_2,S_y=q_2,S_z=
     -\sqrt{1-(S_x^2+S_y^2)}\simeq -1+\frac12 \left(p^2_2+q^2_2\right),
\end{eqnarray*} 
where 
$(q_1,p_1,q_2,p_2)\in T_{\boldsymbol N^*}S^2
\times T_{\boldsymbol S^*}S^2\cong\mathbb R^2\times\mathbb R^2$ 
is a set of local symplectic coordinates. Then 
\begin{eqnarray}
     \fl H_{1/2}(q,p) =
    \underbrace{\textrm{Re}\left[i(q_1-ip_1)(q_2-ip_2)^2\right]}_{H_0}+
      \underbrace{\frac{1}{2}(p_2^2+q_2^2)}_{H_r} 
      -\underbrace{\frac 12(p_1^2+q_1^2)(p_2^2+q_2^2)}_{H_c}, 
        \label{article:localHamiltonian}\\
    \fl J_z(q,p)=-(p_1^2+q_1^2)+\frac 12(p_2^2+q_2^2).
\end{eqnarray}

To find the position of the critical values we solve
\begin{equation}
    DH_{1/2}(q,p)=0, \qquad DJ_z(q,p)=0,
\end{equation}
including terms up to third order $(q_i,p_i<<1)$. There is a corank 
$2$  critical value at $(H,J_z)=(0,0)$ and a line of corank $1$ critical 
points 
\begin{equation}
    (H,J_z)=(0,-p_1^2-q_1^2),
\end{equation} 
in accordance with \fref{article:fig:localMoment}. As $H_r$ only 
depends on $q_2,p_2$ it has no influence on the qualitative picture 
and can be disregarded.

$J_z$ is the Hamiltonian of a pair of oscillators in $1:(-2)$ resonance. 
Together with $H_0$ it is the system of functions in involution used to 
demonstrate the existence of fractional monodromy in \cite{nekoroshev03}. 
 
The third term $H_r$ is positive definite and dominates far from the 
origin assuring compactness of the fibers in a neighborhood of 
$(0,0,0,0)\in\mathbb R^4$. This completes the reduction to normal form 
\cite{nekoroshev03,nekoroshev03b}.


\end{appendix}

\section*{References}

\bibliography{references}

\begin{thebibliography}{10}

\bibitem{CanJPhys}
W.A. Kreiner and A.G. Robiette.
\newblock Measurement and analysis of the $\nu_2$ and $\nu_4$ infrared bands of
  methane-$d_4$.
\newblock {\em Can. J. Phys.}, 57:1969--1981, 1979.

\bibitem{pavlov88}
V.B. Pavlov-Verevkin, D.A. Sadovski\'i, and B.I. Zhilinski\'i.
\newblock On the dynamical meaning of diabolic points.
\newblock {\em Europhys. Lett.}, 6:573--578, 1988.

\bibitem{SIADS}
K.~Efstathiou, D.A. Sadovski\'i, and B.I. Zhilinskii.
\newblock Analysis of rotation-vibration relative equilibria on the example of
  a tetrahedral four atom molecule.
\newblock {\em SIAM J. Appl. Dyn. Syst. (SIADS)}, 3:261--351, 2004.

\bibitem{cushman00}
R.H. Cushman and D.A. Sadovski\'i.
\newblock Monodromy in the hydrogen atom in crossed fields.
\newblock {\em Physica D}, 65:166--196, 2000.

\bibitem{arnold89}
V.I. Arnol'd.
\newblock {\em Mathematical {M}ethods of {C}lassical {M}echanics}.
\newblock Springer, Heidelberg, 1989.

\bibitem{bates97}
R.H. Cushman and L.M. Bates.
\newblock {\em {G}lobal {A}spects of {C}lassical {I}ntegrable {S}ystems}.
\newblock Birkh{\"a}user, Basel, 1997.

\bibitem{sadovskii99}
D.A. Sadovski\'i and B.I. Zhilinski\`i.
\newblock Monodromy, diabolic points, and angular momentum coupling.
\newblock {\em Phys. Lett. A}, 256:235--244, 1999.

\bibitem{faure00}
F.Faure and B.I. Zhilinski\'i.
\newblock Topological {C}hern indices in molecular spectra.
\newblock {\em Phys. Rev. Lett.}, 85:960--963, 2000.

\bibitem{grondin}
L.~Grondin, D.A. Sadovski\'i, and B.I. Zhilinski\'i.
\newblock Monodromy in systems with coupled angular momenta and rearrangement
  of bands in quantum spectra.
\newblock {\em Phys. Rev. A}, 65:012105--1--15, 2002.

\bibitem{duistermaat80}
J.J. Duistermaat.
\newblock On global action angle coordinates.
\newblock {\em Comm. Pur. App. Math.}, 33:687--706, 1980.

\bibitem{duistermaat98}
J.J. Duistermaat.
\newblock The monodromy of the {H}amiltonian {H}opf bifurcation.
\newblock {\em Z. angew. Math. Phys.}, 49:156--161, 1998.

\bibitem{ngoc03}
S.~V{\~u} Ng\d{o}c.
\newblock Moment polytopes for symplectic manifolds with monodromy.
\newblock {\em Adv. Math.}, 208:909--934, 2007.

\bibitem{faure01}
F.~Faure and B.I. Zhilinski\'i.
\newblock Topological properties of the {B}orn-{O}ppenheimer {A}pproximation
  and {I}mplications for the {E}xact {S}pectrum.
\newblock {\em Lett. Mat. Phys.}, 55:219--238, 2001.

\bibitem{nekoroshev03}
N.N. Nekhoroshev, D.A. Sadovski\'i, and B.I. Zhilinski\'i.
\newblock Fractional monodromy of resonant classical and quantum oscillators.
\newblock {\em C. R. Acad. Sci. Paris, Ser.$I$}, 335:985--988, 2002.

\bibitem{nekoroshev03b}
N.N. Nekhoroshev, B.I. Sadovski\'i, and B.I. Zhilinski\'i.
\newblock Fractional {H}amiltonian monodromy.
\newblock {\em Ann. Henri Poincar\'e}, 7:1099--1211, 2006.

\bibitem{KECushSad}
K.~Efstathiou, R.H. Cushman, and D.A. Sadovski\'i.
\newblock Fractional monodromy in the $1:(-2)$ resonance.
\newblock {\em Adv. Math}, 209:241--273, 2007.

\bibitem{KEbook}
K.~Efstathiou.
\newblock Metamorphoses of {H}amiltonian systems with symmetries.
\newblock {\em Lect. Notes Math.}, 1864, 2005.

\bibitem{broer02}
H.~Broer, R.H. Cushman, and F.~Fass\'o.
\newblock Geometry of {KAM} tori for nearly integrable {H}amiltonian systems.
\newblock {\em Ergod. Th \& Dyn. Sys., (arXiv:math.DS/0210043)}, to appear,
  2007.

\bibitem{ngoc99}
S.~V{\~u} Ng\d{o}c.
\newblock Quantum monodromy in integrable systems.
\newblock {\em Commun. Math. Phys.}, 203(2):465--479, 1999.

\bibitem{zung96}
T.Z. Zung.
\newblock Symplectic topology of integrable {H}amiltonian systems {I}:
  {A}rnol'd-{L}iouville with singularities.
\newblock {\em Comp. Math.}, 101:179--215, 1996.

\bibitem{zung02}
T.Z. Zung.
\newblock Symplectic topology of integrable {H}amiltonian systems {II}:
  Topological classification.
\newblock {\em Comp. Math.}, 138:125--156, 2003.

\bibitem{cushman04}
R.H. Cushman, H.R. Dullin, A.~Giacobbe, M.~Joyeux, P.~Lynch, D.A. Sadovski\'i,
  and B.I. Zhilinski\'i.
\newblock {CO}$_2$ molecule as a quantum realization of the $1:1:2$ resonant
  swing-spring with monodromy.
\newblock {\em Phys. Rev. Lett.}, 93:24302, 2004.

\bibitem{joyeux03}
M.~Joyeux, D.A. Sadovski\'i, and J.~Tennyson.
\newblock Monodromy of the ${L}i{NC}/{NCL}i$ molecule.
\newblock {\em Chem. Phys. Lett.}, 382:439--442, 2003.

\bibitem{MolPhys}
D.A. Sadovski\'i and B.I. Zhilinski\`i.
\newblock Quantum monodromy, its generalizations and molecular manifestations.
\newblock {\em Mol. Phys.}, 104:2595--2615, 2006.

\bibitem{child}
M.S. Child.
\newblock Quantum monodromy and molecular spectroscopy.
\newblock {\em Adv. Chem. Phys.}, in press, 2006.

\bibitem{AnnPhys}
D.A. Sadovski\'i and B.I. Zhilinski\`i.
\newblock Hamiltonian systems with detuned $1:1:2$ resonance. manifestations of
  bidromy.
\newblock {\em Ann. Phys. (N.Y.)}, 232:164--200, 2007.

\bibitem{berry84a}
M.V. Berry.
\newblock Quantal phase factors accompanying adiabatic change.
\newblock {\em Proc. R. Soc. Lond. A}, 392:45--57, 1984.

\bibitem{panati03}
G.~Panati, H.~Spohn, and S.~Teufel.
\newblock Space-adiabatic perturbation theory.
\newblock {\em Adv. Theor. Math. Phys.}, 7:145--204, 2003.

\bibitem{faure02a}
F.Faure and B.I. Zhilinski\'i.
\newblock Topologically coupled energy bands.
\newblock {\em Phys. Lett. A}, 302:242--252, 2002.

\bibitem{hansen04}
M.S. Hansen.
\newblock Adiabatically coupled systems: {R}edistribution, monodromy, and
  {C}hern index.
\newblock Master's thesis, Technical University of Denmark, 2004.

\bibitem{zhilinskii01}
B.I. Zhilinski\'i.
\newblock Symmetry, invariants, and topology in molecular models.
\newblock {\em Phys. Rep.}, 341:85--171, 2001.

\bibitem{landau65}
L.~Landau and E.~Lifshitz.
\newblock {\em Quantum {M}echanics (Theo. Phys. Vol. $III$)}.
\newblock Mir, Moscow, 1965.

\bibitem{nakahara90}
M~Nakahara.
\newblock {\em Geometry, {T}opology and {P}hysics}.
\newblock Adam Hilger, New York, 1990.

\bibitem{leboeuf91}
P.~Leb{\oe}uf.
\newblock Phase space approach to quantum dynamics.
\newblock {\em J. Phys. A: Math. Gen.}, 24:4574--4586, 1991.

\bibitem{kurchan89}
J.~Kurchan, P.~Leb\oe uf, and M.~Saraceno.
\newblock Semiclassical approximation in the coherent-state representation.
\newblock {\em Phys. Rev. A}, 40:6800--6813, 1989.

\bibitem{zang90}
W.~Zhang, D.H. Feng, and G.~Gilmore.
\newblock Coherent states: {T}heory and some applications.
\newblock {\em Rev. Mod. Phys.}, 62:867--927, 1990.

\bibitem{michel01}
L~Michel and B~I Zhilinski\'i.
\newblock Symmetry, invariants, topology. {B}asic tools.
\newblock {\em Phys. Rep.}, 341:11--84, 2001.

\bibitem{guillemin}
V.~Guillemin.
\newblock {\em {M}oment {M}aps and {C}ombinatorial {I}nvariants of
  {H}amiltonian $T^n$-spaces}.
\newblock Birkh{\"a}user, Boston, 1994.

\bibitem{marsden99}
J.~E. Marsden and T.~S. Ratiu.
\newblock {\em {I}ntroduction to {M}echanics and {S}ymmetry ($2$nd edition)}.
\newblock Springer, Heidelberg, 1999.

\bibitem{zung97}
T.Z. Zung.
\newblock A note on focus-focus singularities.
\newblock {\em Diff. Geom. Apppl.}, 7:123--130, 1997.

\bibitem{BolFom}
A.V. Bolsinov and A.T. Fomenko.
\newblock {\em {I}ntegrable {H}amiltonian {S}ystems. {G}eometry, {T}opology,
  {C}lassification}.
\newblock Chapman {\&} Hall/CRC, London, 1997.

\bibitem{simon83}
B.~Simon.
\newblock Holonomy, the {Q}uantum {A}diabatic {T}heorem, and {B}erry's {P}hase.
\newblock {\em Phys. Rew. Lett.}, 51:2167--2170, 1983.

\bibitem{cushman88}
R.H. Cushman and J.J. Duistermaat.
\newblock The quantum mechanical spherical pendulum.
\newblock {\em Bull. Am. Soc.}, 19:475--479, 1988.

\bibitem{giacobbe04}
A.~Giacobbe, R.H. Cushman, D.A. Sadovski\'i, and B.I. Zhilinski\'i.
\newblock Monodromy of the quantum $1:1:2$ resonant swing spring.
\newblock {\em J. Math. Phys.}, 45:5076--5100, 2004.

\bibitem{avron88}
J.~Avron and B.~Zur A.~Raveh.
\newblock Adiabatic transport in multiply connected systems.
\newblock {\em Rev. Mod. Phys}, 60:873--915, 1988.

\bibitem{griffiths78}
P.~Griffiths and J.~Harris.
\newblock {\em Principles of algebraic geometry}.
\newblock John Wiley \& Sons, New York, 1978.

\bibitem{fedosov00}
F.~Fedosov.
\newblock The {A}tiyah-{B}ott-{P}atodi {M}ethod in deformation quantizaton.
\newblock {\em Commun. Math. Phys.}, 209:691--728, 2000.

\bibitem{ProcRS}
K.~Efstathiou, D.A. Sadovski\'i, and B.I. Zhilinskii.
\newblock Classification of perturbations of the hydrogen atom by small static
  electric and magnetic fields.
\newblock {\em Proc. Roy. Soc. (London)}, submitted, 2007.

\bibitem{rink04}
B.~Rink.
\newblock A cantor set of tori with monodromy near a focus-focus singularity.
\newblock {\em Nonlinearity}, 17:1--10, 2004.

\end{thebibliography}

\bibliographystyle{unsrt}

\end{document}